\documentclass[12pt]{article}
\pdfoutput=1
\usepackage{amsmath}
\usepackage{amsfonts}
\usepackage{float}
\usepackage{url}
\usepackage{hyperref}
\usepackage{bbold}
\usepackage{slashed}
\usepackage{simplewick}
\usepackage{tikz}
\usepackage{graphicx}
\usepackage{epstopdf}
\usepackage{subfigure}
\usepackage[labelfont=bf,font={small}]{caption}
\usepackage{pgfplots}
\usepackage{amssymb}
\usepackage{color}
\usepackage{mathrsfs}
\usepackage{fancybox}
\usepackage{lipsum}
\usepackage{eurosym}
\usepackage{tcolorbox}
\usepackage{physics}
\usepackage{tensor}
\usepackage{empheq}
\definecolor{someGray}{RGB}{242,242,242}

\newcommand{\dpi}{\mathcal{D}}

\newcommand{\mo}{\mathcal{O}}

\newcommand{\cas}{\mathcal{C}}

\newcommand{\Ai}{\text{Ai}}

\newcommand{\Poincare}{Poincar\'e }
\newcommand{\sinc}{\text{sinc}}
\newcommand{\average}[1]{\left\langle #1 \right\rangle}

		\newcommand{\CFT}{\scriptscriptstyle \text{CFT}}
		
\def\mM{\mbox{\small {\rm M}}}

																\newcommand{\T}{\small \text{T}}

								\newcommand{\vertrule}{\,\rule[-1.2pt]{.2pt}{1.7ex}\,}

\numberwithin{equation}{section}

\begin{document}
\begin{titlepage}

\setcounter{page}{1} \baselineskip=15.5pt \thispagestyle{empty}

\vfil

${}$

\begin{center}
\def\thefootnote{\fnsymbol{footnote}}
\begin{changemargin}{0.05cm}{0.05cm} 
\begin{center}
{\Large \bf Unruh detectors and quantum chaos in JT gravity
}
\end{center} 
\end{changemargin}
\vskip 1cm
{Andreas Blommaert\footnote{\href{mailto:andreas.blommaert@ugent.be}{\protect\path{andreas.blommaert@ugent.be}}}, Thomas G. Mertens\footnote{\href{mailto:thomas.mertens@ugent.be}{\protect\path{thomas.mertens@ugent.be}}} and Henri Verschelde\footnote{\href{mailto:henri.verschelde@ugent.be}{\protect\path{henri.verschelde@ugent.be}}}
}
\\[0.5cm]
{\normalsize { \sl Department of Physics and Astronomy\\
Ghent University, Krijgslaan 281-S9, 9000 Gent, Belgium}}

\vspace{0.2cm}

\medskip

\end{center}


\begin{center} 
\textbf{Abstract}
\end{center} 
We identify the spectral properties of Hawking-Unruh radiation in the eternal black hole at ultra low energies as a probe for the chaotic level statistics of quantum black holes. Level repulsion implies that there are barely Hawking particles with an energy smaller than the level separation. This effect is experimentally accessible by probing the Unruh heat bath with a linear detector. We provide evidence for this effect via explicit and exact calculations in JT gravity building on a radar definition of bulk observables in the model. Similar results are observed for the bath energy density. This universal feature of eternal Hawking radiation should resonate into the evaporating setup.

\vspace{0.3cm}
\vfil
\begin{flushleft}
\today
\end{flushleft}

\end{titlepage}

\newpage
\addtolength{\parskip}{1mm}
\addtolength{\abovedisplayskip}{1mm}
\addtolength{\belowdisplayskip}{1mm}
\tableofcontents
\vspace{0.5cm}
\noindent\makebox[\linewidth]{\rule{\textwidth}{0.4pt}}
\vspace{1cm}

\addtolength{\abovedisplayskip}{.5mm}
\addtolength{\belowdisplayskip}{.5mm}

\def\plus{\raisebox{.5pt}{\tiny$+$\smpc}}

\def\spc{\hspace{1pt}}

\def\nspc{{\hspace{-2pt}}}
\def\ff{\rm\smpc f\smpc} 
\def\fff{\mbox{Y}}
\def\ww{{\rm w}}
\def\smpc{{\hspace{.5pt}}}

\def\zz{{\spc \rm z}}
\def\xx{{\rm x\smpc}}
\def\xxi{\mbox{\footnotesize \spc $\xi$}}
\def\jj{{\rm j}}
\addtolength{\baselineskip}{-.1mm}

\renewcommand{\Large}{\large}

\setcounter{tocdepth}{2}
\addtolength{\baselineskip}{0mm}
\addtolength{\parskip}{0mm}

\setcounter{footnote}{0}

\section{Introduction}\label{sect:1}
One of the main features of finite volume AdS/CFT is that the CFT in question is a unitary and discrete system. The latter follows from the fact that the boundary on which the CFT lives is compact. This implies that quantum gravity in AdS must be a unitary and discrete quantum system. There is however at least one further constraint on the spectrum of quantum gravity in AdS on which we want to focus in this work. It is a consequence of the fact that black holes are chaotic quantum systems \cite{bhrm}. 

Roughly speaking there are two hallmarks of chaotic quantum systems. The first is exponential dependence on changes in initial conditions. In a quantum theory this translates into the exponential growth of operators with time \cite{operatorgrowth}. One popular way to probe this effect is by computing out-of-time-ordered correlators \cite{SS,SSmultiple,Shenker:2014cwa,bound}. The gravitational translation of this fast scrambling is the exponential redshift close to gravitational horizons, resulting in gravitational shockwaves \cite{SS,SSmultiple,Shenker:2014cwa}.

The second hallmark of quantum chaotic systems concerns their level statistics. In particular quantum chaotic systems are very well characterized by the property of level repulsion. Roughly speaking two subsequent energy levels of a chaotic system are rarely close together. A more precise version of this is that the local spectral statistics of any quantum chaotic system can be described using random matrix statistics.\footnote{For a system with internal symmetries, level repulsion or random matrix statistics emerges within each sector with fixed quantum number (representation label), where each energy level is degenerate as dim $R$ with $R$ the representation. See for example \cite{mehta,haake,Kapec:2019ecr}.} 

Black holes are fast scramblers \cite{sekinosusskind} and shockwave interactions are a universal feature of quantum gravity. Therefore quantum black holes are expected to be chaotic quantum systems \cite{bhrm}. But if quantum gravity is quantum chaotic, then what is the bulk gravitational interpretation of level repulsion and of random matrix level statistics? Recently this question has been addressed within Jackiw-Teitelboim (JT) gravity \cite{sss2}.\footnote{Gravitational shockwaves in this model have been studied in \cite{malstanyang,shockwaves}. For another link between JT gravity and quantum chaos, see \cite{Garcia-Garcia:2019zds}.} The answer in this theory is that Euclidean wormhole contributions to the gravitational path integral are responsible for level repulsion and random matrix level statistics in the spectrum of the gravitational theory. We expect this conclusion to be true universally. 
\\~\\
We conclude that any acceptable theory of quantum gravity in AdS is a discrete and unitary quantum system with random matrix level statistics. One prime consequence of this spectral behavior is the specific late time behavior of boundary correlators. 
Consider for example the two-point function of a discrete quantum chaotic system with an $L$ dimensional Hilbert space with levels $\lambda_1\dots \lambda_L$:
\begin{align}
    \average{\mo (0)\mo (t)}_\beta
		&= \int_\cas d E_1\,e^{-\beta E_1}\,\int_\cas d E_2\,e^{it(E_1-E_2)}\,\rho(E_1,E_2)\,\rvert\mo_{E_1 E_2}\rvert^2\,,\label{erraticbdy}
\end{align}
where
\begin{equation}
    \rho(E_1,E_2)=\sum_{i=1}^L \delta(E_1-\lambda_i)\sum_{j=1}^L \delta(E_2-\lambda_j)\,.\label{spikes}
\end{equation}
For a quantum chaotic system the eigenvalue thermalization hypothesis \cite{eth1,eth2,bhrm,phil} states that $\rvert\mo_{E_1 E_2}\rvert^2$ is a smooth function of $E_1$ and $E_2$. At early times the Fourier transform in \eqref{erraticbdy} cannot distinguish the delta functions \eqref{spikes} from a coarse-grained version of this spectrum. The result is that at early times this correlator decays exponentially with time.
This is the quasinormal mode decay known from quantum fields on a black hole background. 

At exponentially late times however the correlator essentially oscillates erratically around an in general nonzero average \cite{maldainfo}. The nonzero averaged value is explained by random matrix theory \cite{bhrm}. The erratic oscillations are testimony to the fundamental discreteness of the theory. In fact this erratically oscillating behavior can itself be viewed as distinguishing a chaotic from a regular quantum system \cite{haake}. 

In gravity, the bulk explanation for the late time behavior of this correlator involves rather exotic gravitational effects. In particular in order to explain the nonzero average one needs to sum over an infinite number of Euclidean wormhole amplitudes in the Euclidean path integral \cite{sss,sss2,Blommaert:2019hjr,phil}. To explain the erratic oscillations in JT gravity from the Euclidean path integral one furthermore needs to realize that a single discrete quantum chaotic system can be described by a version of JT gravity which includes a specific set of branes in the gravitational path integral \cite{paper6,maxfieldmarolf,wophilbert}.
\\~\\
In this work we would like to introduce a structurally different probe of the chaotic level statistics of quantum black holes. We propose to investigate the low-energy spectral properties of the eternal Hawking-Unruh radiation as detected by a linear Unruh-DeWitt detector. There are two main reasons why we believe this to be an interesting observable.
\begin{enumerate}
    \item It has recently been advocated \cite{rw1,rw2} that the same Euclidean wormholes which play a crucial role in explaining the late time behavior of correlators \cite{sss,sss2,Blommaert:2019hjr,phil} are key to understanding unitary black hole evaporation from the bulk point of view.
		Ultimately we want to understand black hole evaporation by tracking what happens to all Hawking quanta emitted from the horizon during the evaporation process,  
		which is more fine-grained bulk information than the early-late entanglement entropy considered in \cite{rw1,rw2}. We expect to already see traces of the mechanism in the eternal Hawking-Unruh radiation. 
    \item Ultimately we want to have a pure bulk gravitational intuition of quantum gravity. In this sense our observable stands out as compared to for example late time boundary correlators. It is inherently a bulk observable. One further incentive to investigate bulk probes in quantum gravity is that in de Sitter or in flat space it is not so natural to formulate questions in the dual theory and so to make progress there we will need to strengthen our understanding of bulk observables in quantum gravity.\footnote{For example in order to probe scrambling in de Sitter one is led to investigate shockwaves and out-of-time-ordered correlators in the bulk \cite{aalsmashiu,Geng:2020kxh,wopjordan}. In this setup we do not have access to intuition from a unitary dual theory \cite{cotleremergent}.}
\end{enumerate}

This work is organized as follows.

In \textbf{section \ref{sect:1}} we argue on general grounds for a modification of the semiclassical formula for the Unruh-Hawking emission probability which accounts for the fact that black holes have chaotic level statistics. We then introduce the setup of JT gravity which we will use to gather concrete evidence for these ideas. 

In \textbf{section \ref{s:udw}} we explicitly verify these expectations. In particular, we couple a massless scalar field to JT gravity and compute the response rate of an Unruh-DeWitt detector which couples linearly to this scalar field. We analyze the detector response in three layers of approximation. Firstly we use the semiclassical approximation. Secondly we discuss the effects of coupling to the Schwarzian reparameterization mode. Finally we include the effects of Euclidean wormholes which give rise to random matrix level statistics. We observe a depletion of the detector response rate at extremely low energies as testimony to the chaotic level statistics of quantum black holes. We generalize to bulk fermionic matter and to other detector couplings.

In \textbf{section \ref{s:hb}} we compute the spectral energy density in the Unruh heat bath. For the Schwarzian system (disk topology) this was investigated in \cite{Mertens:2019bvy}. Perhaps surprisingly this spectral energy density is not precisely identical to the detector response. We explain that this is due to ordering ambiguities that arise when promoting classical expressions to gravitational operators. Despite these subtle differences with the detector setup, we find a similar depletion at ultra low energies due to level repulsion.

In \textbf{section \ref{s:concl}} we comment on gravitational dressings, higher genus contributions to bulk correlators and evaporation. Certain more technical aspects of the story have been relegated to appendices.
\subsection{Unruh detectors and level repulsion}
\label{s:exp}
The main conceptual point in this work is that the semiclassical Planckian black body law for Hawking radiation does not take into account the chaotic level statistics of the quantum black holes which emit these quanta. 
\\~\\
We first briefly review the expected level statistics of black holes. For an arbitrary quantum mechanical system we denote the probability to find an energy level between $E_1$ and $E_1+ d E_1$ and a second energy level between $E_2$ and $E_2+d E_2$ as
\begin{equation}
    \rho(E_1,E_2)\, d E_1\,d E_2.
\end{equation}
For a system of which the level spacings are Poisson distributed, the probability to find a level between $E_1$ and $E_1+d E_1$ is independent of the probability to find a second level between $E_2$ and $E_2+d E_2$
\begin{equation}
    \rho(E_1)\rho(E_2)\, d E_1\, d E_2.
\end{equation}
In other words different levels are uncorrelated. An example of a system with implicit Poisson level statistics is a particle in a very large box, typically used to derive the Planckian black body law. 

For chaotic quantum systems, the story is quite different. The local level statistics of a quantum chaotic system are described by random matrix theory \cite{haake,mehta}. Level statistics in random matrix theory is universal: the multi-density correlators of essentially any chaotic quantum systems (without time-reversal symmetry) are those of the Gaussian unitary ensemble (GUE).\footnote{We will focus on such systems here. The discussion is immediately modified to other ensembles \cite{Stanford:2019vob}.} The GUE two-level correlator is\footnote{There is furthermore a contact term contribution $\rho(E_1)\delta(E_1-E_2)$ which is largely irrelevant to our story here. We define
\begin{equation}
    \sinc\, x =\frac{\sin x}{x}.
\end{equation}}
\begin{equation}
    \rho(E_1,E_2) = \rho(E_1)\rho(E_2)-\rho(E_1)\rho(E_2)\,\sinc^2\, \pi \rho(E_1)(E_1-E_2).\label{re1e2}
\end{equation}
This leads to the normalized two-density correlator
\begin{equation}
    \frac{\rho(E_1,E_2)}{\rho(E_1)\rho(E_2)}=\quad \raisebox{-15mm}{\includegraphics[width=50mm]{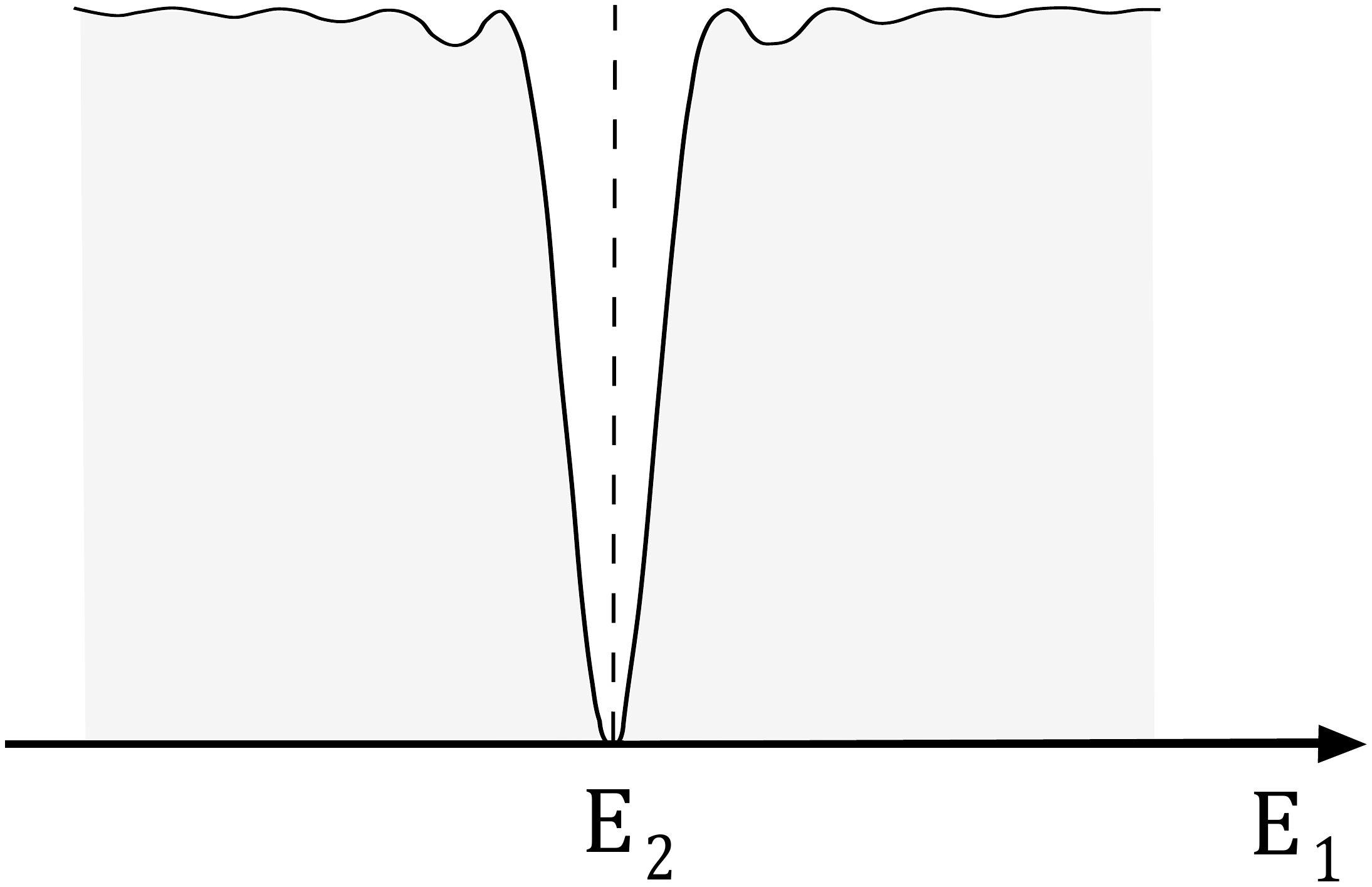}}\quad.\label{8}
\end{equation}
This depletion around $E_1=E_2$ is the characteristic level repulsion of quantum chaotic systems. In particular for the GUE we have quadratic level repulsion meaning there is a quadratic zero in level correlators when any two eigenvalues approach.\footnote{The Gaussian orthogonal ensemble (GOE) and Gaussian symplectic ensemble (GSE) have linear respectively quartic level repulsion.}
\\~\\
Now let's think about a process where a system with such chaotic level statistics emits an energy quantum. In particular let's consider the probability for a black hole with energy $E$ to emit a massless scalar particle with energy $\omega$ which we then detect in our detector. In $d$ spacetime dimensions this probability is proportional to the usual Planckian black body law\footnote{One recognizes the Bose-Einstein distribution. The factor $\omega^{d-1}$ is proportional to the level density for a particle in a $d-1$ dimensional box. This formula should furthermore be augmented with suitable greybody factors associated with the centrifugal barrier and with interference terms associated with reflections off of asymptotic boundaries. We leave them implicit here.}
\begin{equation}
R_{\text{BE}}(\omega)  \sim  \frac{\omega^{d-1}}{e^{\beta \omega}-1}.\label{be}
\end{equation}
The Hawking modes in the Unruh heat bath originate from a decay process between two black hole energy levels. So every energy level $\omega$ of the Hawking radiation corresponds to the difference of two energy levels of the quantum chaotic black hole system
\begin{equation}
    \omega= E_1-E_2.\label{difference}
\end{equation}
Therefore the level density for the Unruh modes in the heat bath is by definition sensitive and proportional to the two-level spectral density $\rho(E_1,E_2)$ of the underlying quantum black hole. 

For this reason we expect that the intrinsically chaotic level statistics of quantum black holes modifies the detection formula \eqref{be} for the probability to detect a massless scalar Hawking particle with energy $\omega$ that has been emitted by a black hole with energy $E$ as
\begin{equation}
\boxed{
     R(\omega) \, \sim\,  R_{\text{BE}}(\omega)(1-\sinc^2 \pi \rho(E)\,\omega\,) \, \sim\,  R_{\text{BE}}(\omega) \frac{\rho(E,E-\omega)}{\rho(E)\rho(E-\omega)}} \,.
		\label{expectationunruh}
\end{equation}
This appropriately takes into account level repulsion.\footnote{The middle formula is the unique rate which probes the two level spectral density for transitions of energy $\omega$ close to a heavy black hole of mass $E$ and which matches with the semiclassical answer \eqref{be} for $\omega\gg 1/\rho(E)$.} Of course this level repulsion is only visible at ultra low energies $\omega$. At such low energies the semiclassical Planckian black body low goes like $\sim \omega^{d-2}$. Level repulsion modifies this behavior at ultra low energies to $\sim \omega^d$. The effect is most clearly visible in 2d where the total detection probability \eqref{expectationunruh} looks like \eqref{8} at very low energies $\omega$.

Before proceeding, let us note that in quantum gravity there will be further modifications to the detector response \eqref{expectationunruh} for highly energetic Hawking modes $\omega\sim E$ for which we may no longer approximate the setup as a light probe particle travelling on a heavy black hole background $E\gg 1$. Related quantum gravitational effects will kick in for Planck size black holes $E\sim 1$. These effects are not governed by random matrix theory.
\subsection{The model}
\label{s:setup}
In the remainder of this work we will gather concrete evidence in favor of \eqref{expectationunruh} via exact calculations in JT gravity. The action of JT gravity is \cite{Jackiw:1984je,Teitelboim:1983ux}\footnote{We work in units where the AdS length $L=1$.}
\begin{equation}
\label{JTaction}
S_{\text{JT}}[g,\Phi] = S_0\,\chi + \frac{1}{16\pi G}\int d x\, \sqrt{g}\,\Phi \left(R + 2\right) + \frac{1}{8\pi G}\int_\partial  d\tau\, \sqrt{h}\,\Phi_{\partial} \left(K- 1 \right)\,.
\end{equation}
Here the Euler character $\chi$ comes from the usual Einstein-Hilbert action in 2d. Its only effect is a weighting of different topologies. The extremal entropy is $S_0 = \Phi_0/4 G$.\footnote{For example if we want to interpret JT gravity as the low-energy limit of a 4d near-extremal black hole, we have $S_0 \sim \frac{Q}{G}$ where $Q$ is the charge of the extremal black hole.} Integrating over the bulk values of the Lagrange multiplier dilaton field $\Phi$ localizes the gravitational path integral on hyperbolic Riemann surfaces $R+2=0$. Any hyperbolic Riemann surface can be built by gluing together different subregions of the \Poincare upper half plane. In Lorentzian signature
\begin{equation}
    ds^2=\frac{d Z^2-d F^2}{Z^2} = -\frac{4d U d V}{(U-V)^2}\quad, \quad Z>0,\label{poinc}
\end{equation}
where we introduced the Poincar\'e lightcone coordinates $U=F+Z$ and $V=F-Z$. There is an asymptotic boundary at $Z=0$ where we will have to impose interesting boundary conditions.

Focusing on one such asymptotic boundary, there are several ways of discussing the reduction of the dynamics of this model into a purely boundary degree of freedom \cite{Almheiri:2014cka,jensen,malstanyang,ads2}. Here we briefly review an intrinsically real-time approach \cite{ads2}. In order to treat this model within the holographic paradigm, we envision a dynamical boundary curve $(F(t),Z(t))$ as UV-cutoff as $Z\to 0$. This curve is specified by off-shell boundary conditions on the metric and dilaton field.

The geometry is taken in Fefferman-Graham gauge, which entails partially fixing the diff-group near the boundary such that the leading part of the geometry is the Poincar\'e metric \eqref{poinc} and the interesting dynamics is in the subleading pieces as $Z \to 0$. Physically this simply means one can only compare different spacetimes if they share the same asymptotics. This leads to the constraint $Z(t) = \varepsilon\, F'(t)$, determining the wiggly boundary in terms of a single reparameterization $F(t)$ mapping the \Poincare time $F$ to the proper time $t$ of an observer following the boundary trajectory. This single function $F(t)$ generates the 1d conformal group (as the residual diff's that preserve Fefferman-Graham gauge) and this is the usual endpoint in AdS/CFT.

However, it can be shown that this system is special in the sense that also the dilaton field $\Phi$ blows up near the boundary $Z=0$. This means it has to be treated on the same footing as the metric, since once again one cannot compare spacetimes with different (dilaton) asymptotics. So, as a second constraint, we choose the boundary curve to satisfy $\Phi_\partial = a/\varepsilon$ in terms of a dimensionful quantity $a$ that determines the theory of interest.\footnote{On-shell, the equations of motion link the dilaton field to the matter content in the theory, and these constraints are then sufficient to find the Schwarzian equation of motion \cite{ads2}.}

Combining the ingredients, one finds that the Lorentzian version of the action \eqref{JTaction} reduces to a Schwarzian derivative action \cite{Almheiri:2014cka,jensen,malstanyang,ads2}
\begin{equation}
\label{SSchL}
S_L[F] = C\int dt \, \text{Sch}(F,t)\quad, \quad \text{Sch}(F,t) =\frac{F'''}{F'} - \frac{3}{2}\left(\frac{F''}{F'}\right)^2.
\end{equation}
The coupling constant $C=a/8\pi G$ has units of length and controls the gravitational fluctuations of the wiggly boundary of the disk.\footnote{It is conventional to choose units such that $C=1/2$.} 
Since path-integral computations are always performed in Euclidean signature, we Wick-rotate $t \to -i\tau$ and $F \to -i F$ to get the Euclidean Schwarzian action
\begin{equation}
\label{SSch}
S[F] =- C\int d\tau \, \text{Sch}(F,\tau)
\end{equation}
With multiple boundaries one has to glue several such Schwarzian systems together \cite{sss2,Blommaert:2018iqz}. Higher genus contributions can be included as explained in \cite{sss2}. In the lowest genus (disk) case it is more convenient to describe the model using the time reparametrization $f(\tau)$ defined as 
\begin{equation}
F(\tau) = \tan \frac{\pi}{\beta} f(\tau)\quad, \quad f(\tau+\beta) = f(\tau) + \beta\quad, \quad \dot{f}(\tau) \geq 0 \, .
\end{equation}
The last two equations characterize $f(\tau)$ as reparametrizing a circle that is the boundary of the 2d Euclidean disk. The quantity $\beta$ is the boundary length and is interpretable as the inverse temperature. We will adhere to this notation of $F(\tau)$ and $f(\tau)$ in the remainder of this work.
\\~\\
Using a myriad of techniques, JT gravity has been exactly solved for an entire class of boundary correlation functions. Gravitational contributions to correlation functions in JT gravity come in several flavors in terms of $G$. 
\begin{enumerate}
    \item There are perturbative $G$ corrections. These can be viewed as boundary graviton interactions and can be obtained via Schwarzian perturbation theory \cite{malstanyang,Stanford:2017thb,Qi:2019gny}. 
    \item There are nonperturbative $G$ corrections associated with an exact solution of Schwarzian correlation functions \cite{altland,altland2,schwarzian,Mertens:2018fds,paper3,Blommaert:2018iqz,kitaevsuh,zhenbin,Iliesiu:2019xuh,suh}. 
    \item There are furthermore nonperturbative $G$ corrections associated with Euclidean wormhole contributions to the Euclidean path integral \cite{sss,sss2,Blommaert:2018iqz,phil,wophilbert}. These represent higher genus Riemann surfaces ending on the wiggly boundary. Indeed via \eqref{JTaction} we see that such contributions are perturbative in $e^{S_0}$ and hence nonperturbative in $G$.
    \item Finally there are nonperturbative $e^{S_0}$ contributions which are hence doubly nonperturbative in $G$. In the gravitational language these are due to brane effects. JT gravity has a dual formulation as a double-scaled random matrix model \cite{sss2}. The doubly nonperturbative effects can be considered as hallmarks of this dual matrix integral description. 
\end{enumerate}
One of the properties which makes JT gravity so interesting is that we have analytic control over all these types of corrections. The centerpiece formulas \eqref{51} and \eqref{323} take all such corrections into account. 


\section{Detectors in the Unruh heat bath}
\label{s:udw}
In this sector we will consider a massless scalar field minimally coupled to JT gravity:\footnote{For earlier discussion in this context see for example \cite{Blommaert:2019hjr,Mertens:2019bvy}.}
\begin{equation}
\label{fieldact}
S= \frac{1}{2} \int d x\,\sqrt{-g}\, g^{\mu\nu}\,\partial_\mu \phi\, \partial_\nu \phi.
\end{equation}
In particular, we aim to probe the emission spectrum of massless scalar Hawking-Unruh particles by a 2d quantum black hole. To do so, we imagine an experiment where we probe the heat bath using a linear Unruh-DeWitt detector. We will isolate the effects of different types of gravitational interactions by working in three improving layers of approximation.
\subsection{Semiclassical analysis}
Let us first consider physics on the gravitational saddle, which is a black hole with inverse Hawking temperature $\beta$
\begin{equation}
    ds^2 = \frac{\pi^2}{\beta^2}\frac{ dz^2-dt^2}{\sinh^2 \frac{2\pi}{\beta} z}\quad, \quad z>0\,.\label{thermal}
\end{equation}
The semiclassical relation between the ADM mass $M$ of the black hole and the Hawking temperature is $\sqrt{M}=2\pi C/\beta$. We will henceforth set $C=1/2$. The asymptotic boundary is at $z=0$ and the semiclassical horizon is at $z=\infty$. This is just a conformal rescaling of flat space. The massless field $\phi$ is insensitive to this conformal rescaling of the metric and hence the solutions to the equations of motion of \eqref{fieldact} are left-and right-moving plane waves. Introducing lightcone coordinates $u=t+z$ and $v=t-z$ we have the mode expansion
\begin{equation}
\label{modex}
\phi(u,v) = \int_0^\infty \frac{d\omega}{\sqrt{4\pi \omega}}\,\left( a_\omega e^{-i\omega u } + a^\dagger_\omega e^{i\omega u }  - a_\omega e^{-i\omega v } - a^\dagger_\omega e^{i\omega v }\right),
\end{equation}
implementing Dirichlet boundary conditions on the boundary $z=0$ of the half plane \eqref{thermal}. The modes are orthogonal with respect to the usual Klein-Gordon inner product, and the modes satisfy
\begin{equation}
    [a_\omega, a^{\dagger}_{\omega'}] = \delta(\omega-\omega').\label{24}
\end{equation}
The Wightman two-point function in the thermal state of the CFT is \cite{spradlin}
\begin{equation}
\average{\phi(u_1,v_1)\phi(u_2,v_2)}_{\CFT} = - \frac{1}{4\pi} \ln \abs{\frac{\sinh \frac{\pi}{\beta}(u_1-u_2)\sinh \frac{\pi}{\beta}(v_1-v_2)}{\sinh \frac{\pi}{\beta}(u_1-v_2)\sinh \frac{\pi}{\beta}(v_1-u_2)}}.\label{25}
\end{equation}
All matter correlators will be denoted by $\average{\dots}_{\CFT}$. This expression can be equivalently read as evaluated in the Poincar\'e vacuum state, defined by taking positive-frequency modes with respect to the \Poincare time $F$. As is well known the correlator looks thermal in $t$ coordinates because of the thermal coordinate transformation $F = \tanh \frac{\pi}{\beta} t$ relating the two \cite{spradlin}. We will be interested in understanding the frequency content of this correlator, and in particular on what it has to say about the underlying black hole.
\\~\\
The Unruh-DeWitt detector is a simple quantum mechanical detector model \cite{Unruh:1976db,DeWitt:1980hx}. It linearly couples a quantum mechanical system with degree of freedom $\mu(t)$ to the scalar quantum field $\phi(u(t),v(t))$ via the local interaction Hamiltonian:
\begin{equation}
    H_\text{int}(t) = g \, \mu(t)\, \phi(u(t),v(t)).\label{26}
\end{equation}
Typically one models the detector system to be a two-level system described by its ``monopole moment'' $\mu$. Here $g \ll 1$ is a tiny coupling and $(u(t),v(t))$ is the worldline of the detector. We assume the detector is initially in its ground state $\ket{0}$. 

We want to compute the probability $P(\omega)$ for the final state of the detector to be the energy eigenstate $\ket{\omega}$ in first order perturbation theory in the detector coupling $g$. Within perturbation theory, the Hilbert space factorizes as  $\mathcal{H}_{\text{det}} \otimes \mathcal{H}_{\text{matter}}$. Since the detector is ignorant of the final state $\ket{\psi}$ of the matter sector, one finds:
\begin{equation}
\label{27}
P(\omega) =\sum_{\psi} \left|\bra{0,M} -i\int_{-\infty}^{+\infty} d t H_\text{int}(t)\ket{\omega,\psi}\right|^2.
\end{equation}
Here $\ket{M}$ denotes the thermal state of the matter sector. Summing over $\psi$, one finds the response rate  $R(\omega)$ which is defined as the probability per unit time to see the detector transition:\footnote{The Fourier transform appears due to
\begin{equation}
    \bra{\omega}\mu(t)\ket{0}=e^{i\omega t}\bra{\omega}\mu(0)\ket{0}.
\end{equation}}
\begin{align}
\label{29} R(\omega)=& \, g^2\,\abs{\bra{\omega}\mu(0)\ket{0}}^2\\&\lim_{{\T} \to +\infty}\frac{1}{\T} \int_{-\T}^{+\T}dt_1\int_{-\T}^{+\T}dt_2\, e^{-i\omega (t_1-t_2)}\average{\phi(u(t_1),v(t_1))\phi(u(t_2),v(t_2))}_{\CFT}.\nonumber
\end{align}
This quantity represents the probability rate for the detector to get excited to energy level $\omega$. By energy conservation the matter state gets depleted by a similar energy $\omega$. 

In the vacuum associated to the time coordinate $t$ the result would be zero but this is not so in the Poincar\'e vacuum. This is the essence of the Unruh effect. The analogous situation in flat space is by interpreting the black hole time coordinate $t$ as Rindler (or Schwarzschild) time, and the Poincar\'e time $T$ as Minkowski time. For a thorough recent review see \cite{Crispino:2007eb}. In the \Poincare vacuum one finds the semi-classical thermal Unruh population \eqref{be} by computing the integral on the second line of \eqref{29} using \eqref{25} and assuming a stationary detector worldline. The answer is
\begin{equation}
		\label{210}
		R(\omega)  = g^2 \abs{\bra{\omega}\mu(0)\ket{0}}^2 \, 2\, \frac{\sin^2 \omega z}{\omega^2}\, \frac{\omega}{e^{\beta \omega} -1}\,.
\end{equation}
One indeed recognizes the Planckian black body law in two dimensions \eqref{be}. 

\subsection{Coupling to Schwarzian reparameterizations}
Our goal for this subsection and the following is to compute the Fourier transform of the bulk two-point function on the second line of \eqref{29} in two different levels of approximation. In this subsection we will ignore Euclidean wormhole (or higher genus) contributions to the Euclidean JT gravity path integral. This means we incorporate the gravitational corrections of only items 1 and 2 of the list in section \ref{s:setup}. 
\\~\\
Within quantum gravity, physical bulk locations and bulk observables must be defined in a diff-invariant manner. 
To do so in a holographic context we are led to define a point in the bulk by anchoring the bulk point to the asymptotic boundary \cite{Donnelly:2015hta,gid3,Almheiri:2017fbd,ref3,Lewkowycz:2016ukf,ref4,ref5,ref6,Chen:2017dnl,Chen:2018qzm,Engelhardt:2016wgb}. One particularly natural way to do this in this context is by geodesic localizing using lightrays \cite{Engelhardt:2016wgb,Blommaert:2019hjr,Mertens:2019bvy}. 

In JT gravity the boundary is one-dimensional. Therefore the physical coordinates used to define bulk points are boundary time coordinates. In particular we need two such time coordinates $t_1$ and $t_2$. We associate these to the lightcone coordinates $u$ and $v$ of a point in the bulk. Here $v$ is the physical boundary time at which an observer sends a signal to a given bulk point and $u$ is the physical boundary time at which the observer receives back the signal after reflecting off some fictitious mirror. 

The boundary curve $(F(t),Z(t))$ to which we anchor a bulk point is described by a single function $F(t)$, where $Z$ is determined in terms of $F$ by the boundary conditions. The actual wiggling of the boundary as explained above \eqref{SSch}, is negligible for $\varepsilon\ll 1$. The field $F(t)$ represents the map between \Poincare coordinates and the physical boundary time coordinate $t$ \cite{Almheiri:2014cka,jensen,malstanyang,ads2}. Consequently in terms of the two boundary times $u$ and $v$, the location of the bulk point in \Poincare coordinates is defined as $U= F(u)$ and $V =F(v)$. For a more detailed explanation of this construction, see \cite{Blommaert:2019hjr,Mertens:2019bvy,wopjordan}. The bulk metric constructed in this fashion is
\begin{equation}
    ds^2=-\frac{F'(u)F'(v)}{(F(u)-F(v))^2}\,du\,dv.\label{189}
\end{equation}
Following this same logic, we are led to define massless scalar bulk observables $\Phi$ which implicitly depend on the Schwarzian reparameterization $f$ as:
\begin{equation}
\label{dress}
\Phi[f\vertrule u,v] = \phi(f(u),f(v)).
\end{equation}
This is just a regular massless scalar field but the location of the insertion of this operator in \Poincare coordinates $(Z,F)$ depends on the details of the wiggly boundary $F(t)$
\begin{equation}
    \Phi[f_1\vertrule u,v]= \raisebox{-17mm}{\includegraphics[width=21mm]{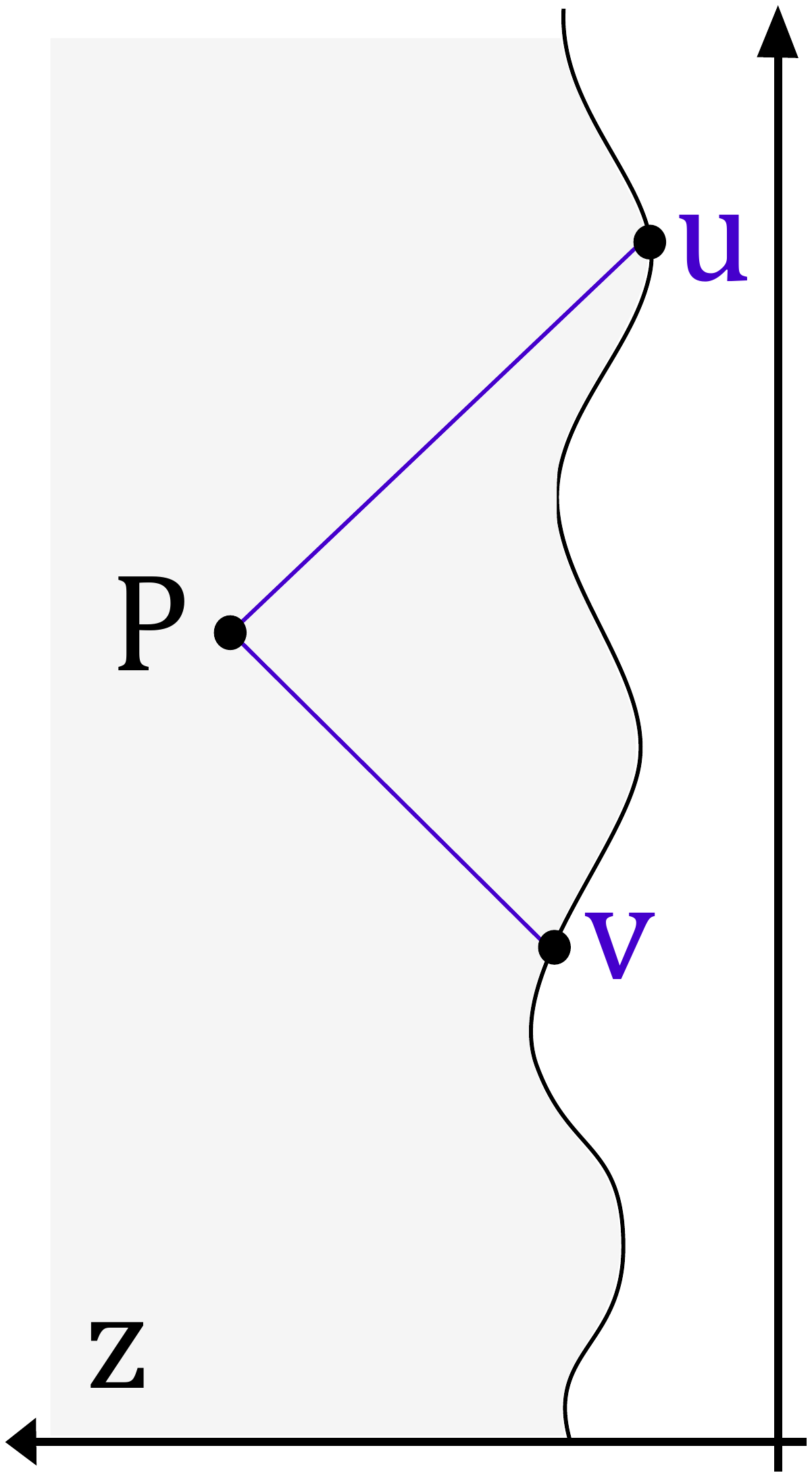}}\quad,\quad \Phi[f_2\vertrule u,v]= \raisebox{-17mm}{\includegraphics[width=22.7mm]{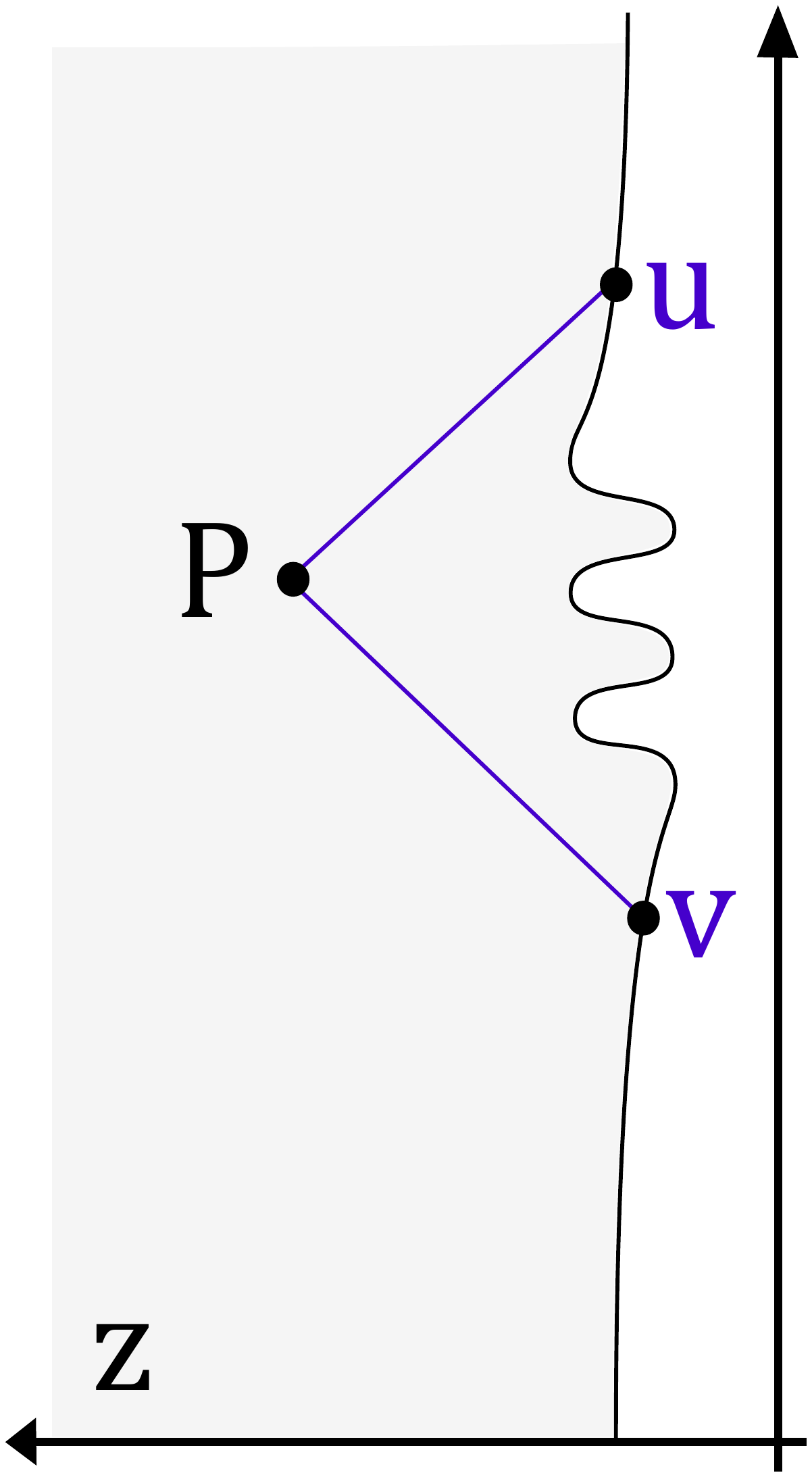}}\quad.
\end{equation}
This implicit dependence on $F(t)$ of bulk operators couples bulk matter to the Schwarzian reparameterization mode. We note that the definition \eqref{dress} of bulk operators in quantum gravity corresponds in the language of \cite{Donnelly:2015hta,gid1,gid2,gid3,gid4,gid5,gid6,gid7} to specifying a particular gravitational dressing of a bare bulk matter operator such that the dressed operator is diff-invariant.
\\~\\
In every fixed metric \eqref{189}, the two-point function of two operators of the type \eqref{dress} is then by definition of \eqref{189} just the reparameterization of \eqref{25}
\begin{equation}
\label{twoqft}
    \average{\Phi[f\vertrule u_1,v_1] \Phi[f\vertrule u_2,v_2]}_{\CFT} = - \frac{1}{4\pi}\left|\ln \frac{\sinh \frac{\pi}{\beta}(f(u_1)-f(u_2))\sinh \frac{\pi}{\beta}(f(v_1)-f(v_2))}{\sinh \frac{\pi}{\beta}(f(u_1)-f(v_2))\sinh \frac{\pi}{\beta}(f(v_1)-f(u_2))}\right|.
\end{equation}
In this same way we define the entire trajectory of the Unruh-DeWitt detector:\footnote{Infinitesimally separated bulk points along such a worldline are separated by
\begin{equation}
ds^2 = -\frac{F'(t+z)F'(t-z)}{(F(t+z)-F(t-z))^2}dt^2
\end{equation}
and since $F' \geq 0$ are hence for any off-shell $F$ time-like separated, proving that the resulting trajectory is always timelike. It is possible for the trajectory to become lightlike at points where time stops flowing $F'=0$.}
\begin{equation}
\includegraphics[width=0.2\textwidth]{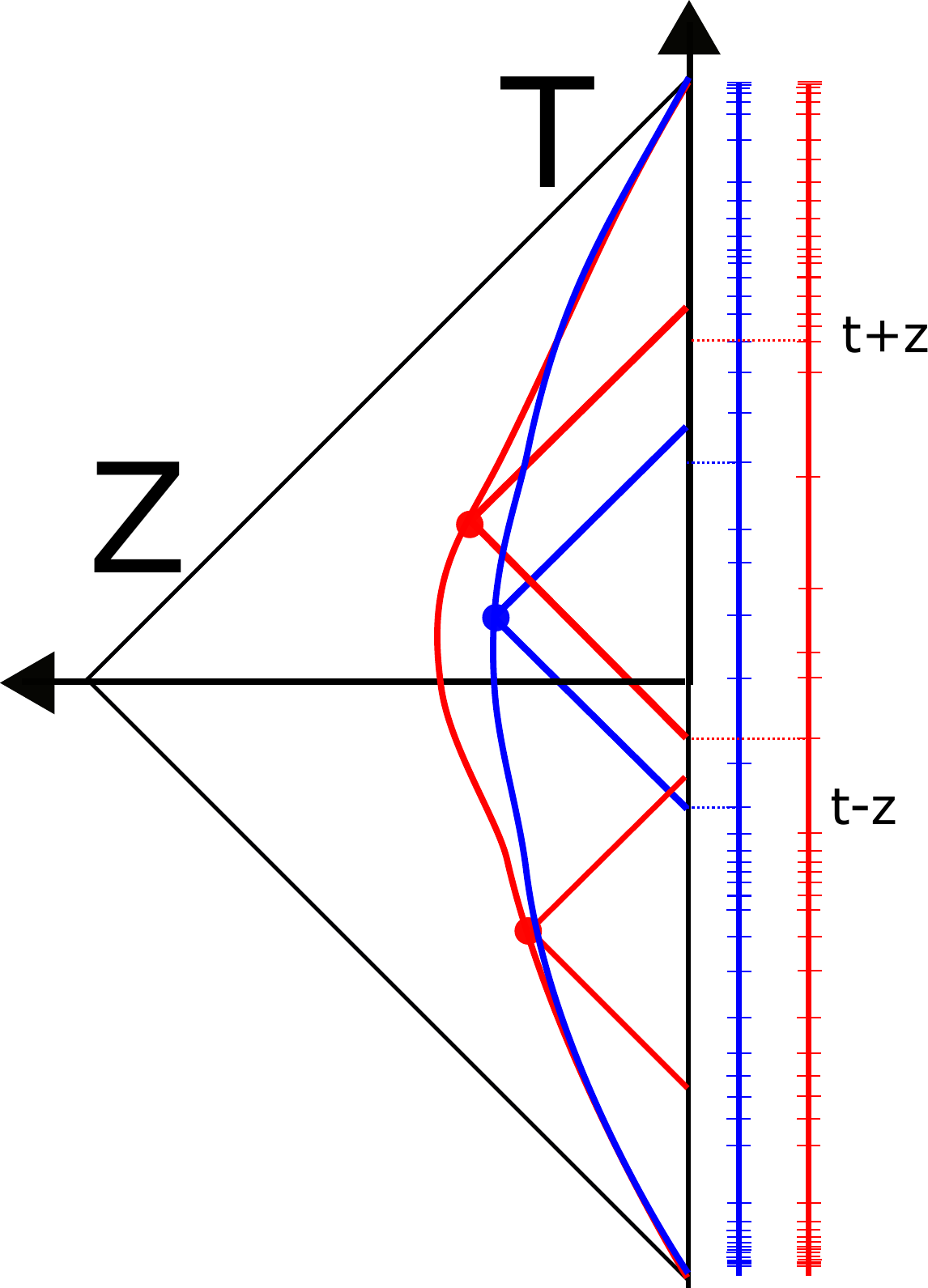}
\end{equation}
This shows the constant $z$ worldlines of the Unruh-DeWitt detector for two different clock ticking patterns $F(t)$ (blue and red) at the same radial coordinate $z$.
Close to the horizon and semi-classically, these become constant accelerated worldlines. Close to the boundary at $z=\epsilon$, this becomes the wiggly boundary curve itself.

The interaction Hamiltonian \eqref{26} now features the diff-invariant dressed field $\Phi$
\begin{equation}
H_\text{int}(t) = g\, \mu(t)\,\Phi[f\vertrule u(t),v(t)].\label{216}
\end{equation}
Including gravitational degrees of freedom and perturbatively in $g$ the Hilbert space may be considered to factorize as $\mathcal{H}_\text{det}\otimes \mathcal{H}_\text{matter+gravity}$. The second factor is the whole coupled system of matter and gravity. Computing the probability for the detector to evolve from an initial state $\ket{0}$ to a final state $\ket{\omega}$ one finds an identical formula \eqref{27} but with the interaction replaced by \eqref{216} and with $\psi$ summed over $\mathcal{H}_\text{matter+gravity}$. Doing the sum over the final states $\psi$ results in
\begin{align}
\bra{\mM}\Phi[f\vertrule u_1,v_1]\Phi[f\vertrule u_2,v_2]\ket{\mM} =\int_{\text{micro M}} [\dpi f]\, \average{\Phi[f\vertrule u_1,v_1] \Phi[f\vertrule u_2,v_2]}_{\CFT}\,e^{-S[f]}.\label{215}
\end{align}
This implements a matter-coupled quantum gravity path integral in the microcanonical ensemble of fixed energy $M$, defined by inverse Laplace transforming the canonical ensemble path integral. Here the Schwarzian action is \eqref{SSch}.
\\~\\
We can now proceed with the actual computation, for which we use a technical trick \cite{Blommaert:2019hjr}:
\begin{equation}
    \average{\Phi[f\vertrule u_1,v_1] \Phi[f\vertrule u_2,v_2]}_{\CFT}=\int_{v_1}^{u_1} d t_1\int_{v_2}^{u_2}dt_2\,\average{\mo[f \vertrule t_1]\mo[f\vertrule t_2]}_{\CFT}.\label{hkll}
\end{equation}
Here the matrix element on the right hand side is a thermal boundary two-point function of a massless scalar:
\begin{equation}
 		    \average{\mo[f\vertrule t_1]\mo[f\vertrule t_2]}_{\CFT}=-\frac{1}{4\pi}\,\frac{ f'(t_1)f'(t_2)}{\frac{\beta^2}{\pi^2}\sinh^2 \frac{\pi}{\beta}(f(t_1)-f(t_2))}.\label{218}
\end{equation}
We notice that \eqref{hkll} is essentially implementing the HKLL bulk reconstruction in each of the metrics \eqref{189} \cite{hkll1,hkll2,kll,kl,Lowe:2008ra}. This reverse-engineered version of bulk reconstruction is pivotal since the Schwarzian path integral in \eqref{215} over the right hand side of \eqref{hkll} can be easily computed \cite{Blommaert:2019hjr}, as the disk boundary-to-boundary propagator in JT gravity:
\begin{equation}
    \int [\dpi f]\, \average{\mo[f\vertrule t_1]\mo[f\vertrule t_2]}_{\CFT}\, e^{-S[f]}\,=\quad\raisebox{-10mm}{\includegraphics[width=45mm]{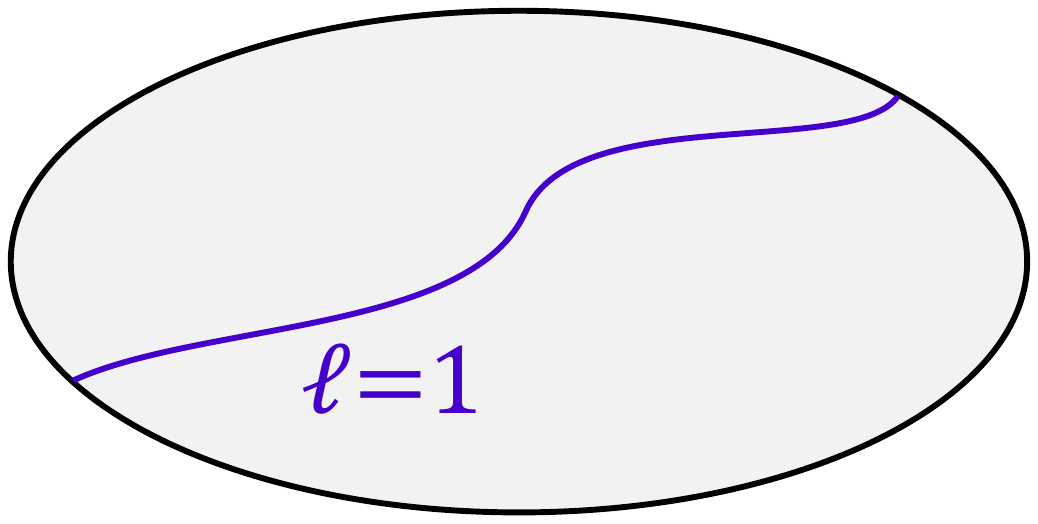}}\quad.\label{40}
\end{equation}
The answer is \cite{altland,altland2,schwarzian,Mertens:2018fds,paper3,Blommaert:2018iqz,kitaevsuh,zhenbin,Iliesiu:2019xuh}:\footnote{There is an implicit $i\varepsilon$ in the exponential $t_1-t_2-i\varepsilon$ as a Euclidean damping factor \cite{schwarzian}. This is related to the particular ordering of both operators in the Wightman two-point function. We will be more explicit about this in section \ref{s:hb} where both operator orderings are relevant.}
\begin{equation}
		    \frac{1}{Z(\beta)}\int_0^\infty d M\,e^{-\beta M}\int_{0}^{+\infty} dE \, e^{-i(t_1-t_2)(E-M)}\, \rho_0(M)\rho_0(E)\,\rvert \mo_{M E}^1\rvert^2. \label{41}
\end{equation}
Here we introduced the notation:\footnote{One should multiply all signs.}
\begin{equation}
	\label{221}
		    \rho_0(E)=\frac{e^{S_0}}{2\pi^2}\sinh 2\pi \sqrt{E}\quad,\quad \rvert\mo_{E_1E_2}^\ell\rvert^2 = e^{-S_0}\frac{\Gamma(\ell\pm i\sqrt{E_1}\pm i\sqrt{E_2})}{\Gamma(2\ell)}\, .
\end{equation}
Within the microcanonical ensemble of fixed energy $M$, the boundary-to-boundary correlator becomes:
\begin{equation}
    \int_{0}^{+\infty} d E \, e^{-i(t_1-t_2)(E-M)}\, \rho_0(E)\,\rvert \mo_{M E}^1\rvert^2 \, .
\end{equation}
The response rate for a detector at $z_1=z_2=z$ is now readily found by performing the elementary integrations in \eqref{29} and \eqref{hkll}. One obtains:
\begin{align}
\label{response}
R(\omega) &= g^2\,\abs{\bra{\omega}\mu(0)\ket{0}}^2\, 2 \, \frac{\sin^2 \omega z}{\omega^2}\, \rho_0(M-\omega)\, \rvert \mo_{M M-\omega}^1\rvert^2 \,.
\end{align}
Notice that by \eqref{221} it is independent of the ground-state degeneracy $e^{S_0}$. The ``greybody'' interference factor $2 \sin^2 \omega z/\omega^2$ is determined in part by choosing Dirichlet boundary conditions for the massless scalar field in \eqref{25}.\footnote{Note that for $z\ll 1$ this quantity goes to zero like $z^2$ as demanded by the extrapolate dictionary for a massless scalar field.} For other boundary conditions on the scalar field, the interference pattern changes, see appendix \ref{app:obc}. 

In the semiclassical regime $M \gg 1$ and $\omega\ll M$, \eqref{response} reproduces the classical answer \eqref{210}. For more generic values of $\omega$ however there are gravitational backreaction effects. The most prominent such backreaction effect is that the response function \eqref{response} abruptly stops at $\omega =  M$ with a square root edge via \eqref{221}.
\\~\\
In order to better understand from a physical point of view why \eqref{response} vanishes at $\omega=M$, it is convenient to go back to the mode expansions \eqref{modex}. We can use it to define raising and lowering operators in the matter Hilbert space:
\begin{align}
\label{creaan}
a_\omega &= \frac{i}{\sqrt{\pi \omega}} \int_{-\infty}^{+\infty} du \,\partial_u \phi(u) \, e^{i\omega u}, \qquad a^{\dagger}_\omega = -\frac{i}{\sqrt{\pi \omega}} \int_{-\infty}^{+\infty} du \, \partial_u \phi(u) \, e^{-i\omega u} \, .
\end{align}
Quite analogously, we could define raising and lowering operators of gravitationally dressed matter fields \eqref{dress} in the Hilbert space of the coupled system of matter and gravity
\begin{align}
\label{creabn}
A_\omega[f] &= \frac{i}{\sqrt{\pi \omega}} \int_{-\infty}^{+\infty} du \,\partial_u \Phi[f\vertrule u] \, e^{i\omega u}, \qquad A^{\dagger}_\omega[f] = -\frac{i}{\sqrt{\pi \omega}} \int_{-\infty}^{+\infty} du \, \partial_u \Phi[f\vertrule u] \, e^{-i\omega u}\, .
\end{align}
Working with these dressed (or diff-invariant) raising and lowering operators, we find that the response rate \eqref{response} is proportional to a number operator expectation value\footnote{Here the factor $V$ is the total volume outside the semiclassical black hole horizon and is an artifact of the matter theory.}
\begin{align}
\label{num1}
\bra{\mM}A^\dagger_\omega[f] A_\omega[f] \ket{\mM} &= \frac{V}{2\pi\omega} \,\rho_0(M-\omega)\, \rvert\mathcal{O}_{M M-\omega}^1\rvert^2\,.
\end{align}
This formula explains the step function in a quite natural manner: the dressed operator $A_\omega$ extracts an energy $\omega$ from the gravity system. Of course it is impossible to extract more energy from this system than the finite energy $M$ which it had to begin with and so we have
\begin{equation}
    A_\omega[f] \ket{\mM}= 0\quad,\quad \omega>M.\label{229}
\end{equation}
Therefore \eqref{num1} and \eqref{response} should be expected to vanish for $\omega >M$. Using similar techniques, one could compute more involved matrix elements. For example
\begin{equation}
    \bra{\mM}A^\dagger_{\omega_1}[f]\dots A^\dagger_{\omega_n}[f]A_{\omega_n}[f]\dots A_{\omega_1}[f]\ket{\mM}.
\end{equation}
One finds, in accordance with the fact that operators such as $A_\omega$ deplete the system of an energy $\omega$, that this amplitude vanishes if $\omega_1+\dots \omega_n>M$. 
\subsection{Level repulsion}
\label{s:levrep}
We would now like to include Euclidean wormhole contributions to the JT gravity path integral which computes the massless bulk two point function. This includes all 4 items of the list of gravitational corrections in section \ref{s:setup}.
\\~\\
It is not a priori obvious how to dress the massless scalar bulk two-point function, whose radar construction was inherently Lorentzian, with higher genus contributions to the Euclidean path integral. We here propose a very natural way of doing so.

We start with the disk (genus zero) contribution first. One may then use the bulk reconstruction formula \eqref{hkll} to write the bulk matter two-point function in terms of a Euclidean disk JT gravity path integral with a boundary-to-boundary matter propagator
\begin{equation}
   \bra{\mM}\Phi[f\vertrule u_1,v_1]\Phi[f\vertrule u_2,v_2]\ket{\mM}\,=\int_{v_1}^{u_1}dt_1\int_{v_2}^{u_2} dt_2 \quad\raisebox{-10mm}{\includegraphics[width=45mm]{1lbiloc.pdf}}\quad.\label{231}
\end{equation}
Including Euclidean wormhole connections for the boundary two-point function on the right hand side has recently been understood \cite{phil,Blommaert:2019hjr,wophilbert}. One just sums over all higher genus Riemann surfaces which end on the union of the boundary circle and the boundary-to-boundary bilocal line. 

We now define bulk correlators by applying the bulk reconstruction formula to these higher-topology boundary correlators. So our definition of bulk operators is really a ``bulk reconstruction first'' approach. This includes for example a contribution of the type
\begin{equation}
    \bra{\mM}\Phi[f\vertrule u_1,v_1]\Phi[f\vertrule u_2,v_2]\ket{\mM}\,\supset\int_{v_1}^{u_1}dt_1\int_{v_1}^{u_2} dt_2 \quad\raisebox{-10mm}{\includegraphics[width=45mm]{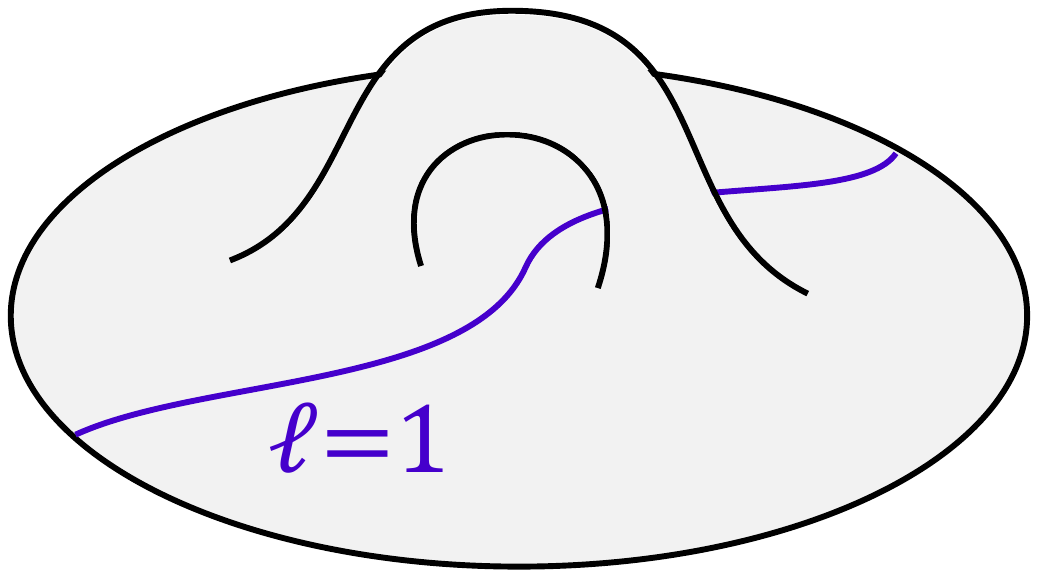}}\quad.\label{232}
\end{equation}
This definition ends up reproducing our generic intuition \eqref{expectationunruh} which we view as an argument that this definition makes sense. 

As it turns out, it is quite feasible to sum over all such amplitudes via the definition of JT gravity as a double-scaled matrix integral \cite{sss2}. In the end, the effect of including such contributions is quite elegant. In terms of the formulas, one ends up effectively replacing $\rho_0(M)\rho_0(E)$ in the boundary two-point function \eqref{41} by $\rho(M,E)$ which is the universal answer \eqref{re1e2} from random matrix theory\footnote{The integration contour is chosen as in \cite{sss2}. It follows the positive real axis. However in the classically forbidden region $E<0$ one needs to choose an appropriate contour for convergence. The contributions to observables from the forbidden part of the integration contour is highly suppressed in all cases by powers of $e^{-S_0}$. Therefor we can essentially neglect these contributions.}
\begin{equation}
		    \frac{1}{Z(\beta)}\int_\cas d M\,e^{-\beta M}\int_\cas dE \, e^{-i(t_1-t_2)(E-M)}\, \rho(M,E)\,\rvert \mo_{M E}^1\rvert^2. \label{233}
\end{equation}
Here we have
\begin{equation}
\rho(M,E) =  \rho(M)\rho(E)-\rho(M)\rho(E)\,\sinc^2\, \pi \rho(M)(M-E) + \rho(M)\delta(M-E). \label{49}
\end{equation}
This formula holds if both $M$ and $E$ are far enough from the spectral edge $E\gg e^{-2S_0/3}$. Closer to the spectral edge, we can resort to similarly universal formulas for the two-level spectral density in the Airy model \cite{sss2,paper6}. The first term corresponds geometrically to the disconnected contribution of adding higher topology to each of the two sides of the bilocal line in \eqref{231}. The second term is due to Riemann surfaces connecting both sides of the bilocal such as the annulus in \eqref{232}. Each of these terms furthermore includes nonperturbative contributions in $e^{-S_0}$ due to brane effects \cite{sss2,phil,paper6}. These are oscillatory and hence not necessarily small. The final term represents a contact term that may or may not have a geometric interpretation.\footnote{We note \cite{phil} that \eqref{233} is conform the idea that this expression represents the ensemble average over different Hamiltonians $H$ of the two point function in a discrete quantum chaotic system. The idea is to take such a set of discrete quantum chaotic systems and to first ensemble average over unitaries $U$ which diagonalize $H$. One then invokes a version of the eigenvalue thermalization hypothesis
\begin{equation}
    \sum_{a,b}\mo_{a}\mo_{b}\int d U\,U_{i a}\,U_{j a}^*\,U_{j b}\,U_{i b}^* = \rvert\mo_{E_i E_j}\rvert^2.
\end{equation}
The assumption is that one point functions in the averaged theory vanish (which they do in JT gravity). Furthermore $\rvert\mo_{E_i E_j}\rvert^2$ are smooth functions on energy scales of order the typical level spacing $e^{-S_0}$. The result is \eqref{erraticbdy}. Furthermore ensemble averaging over the eigenvalues of the Hamiltonians (with a well chosen potential $V(H)$ for Hamiltonians $H$ in the ensemble average) one then indeed finds \eqref{233}.}

We can now immediately write down the analogue of \eqref{response} which takes into account these Euclidean wormholes via the substitution \eqref{233}
\begin{equation}
     \boxed{R(\omega) = g^2\,\abs{\bra{\omega}\mu(0)\ket{0}}^2\,2\,\frac{\sin ^2\omega z}{\omega^2}\,  \frac{\rho(M,M-\omega)}{\rho(M)}\, \rvert \mo_{M M-\omega}^1\rvert^2.}\label{51}
\end{equation}
It is straightforward to similarly write down the modifications to the expectation values of the dressed number operator \eqref{num1}. Notice that now there is no step function. This is because in the matrix integral description there is a tiny but nonzero probability for eigenvalues to be in the forbidden region $E<0$ of the energy contour $\cas$ \cite{sss2}. 

We note that the prefactor $g^2\,\abs{\bra{\omega}\mu(0)\ket{0}}^2$ is intrinsic to the detector, which we (the observer) can determine within the free detector theory. 
Dividing out this known prefactor, and computing \eqref{51} in the probe approximation $M\gg 1$ and $\omega\ll M$ we find
\begin{equation}
    \frac{R(\omega)}{2 \, \sin^2 \omega z/\omega^2} = \frac{\omega}{e^{\beta\omega}-1} (1-\sinc^2 \pi \rho(M)\,\omega\,) \, . \label{237}
\end{equation}
This explicitly confirms our general expectation \eqref{expectationunruh} about the effects of level repulsion on the detection rate, via a bulk JT gravity calculation. Schematically the detector finds the following result
\begin{equation}
	    R(\omega) 	= 2 \, \frac{\sin^2 \omega z}{\omega^2}\quad \raisebox{-15mm}{\includegraphics[width=65mm]{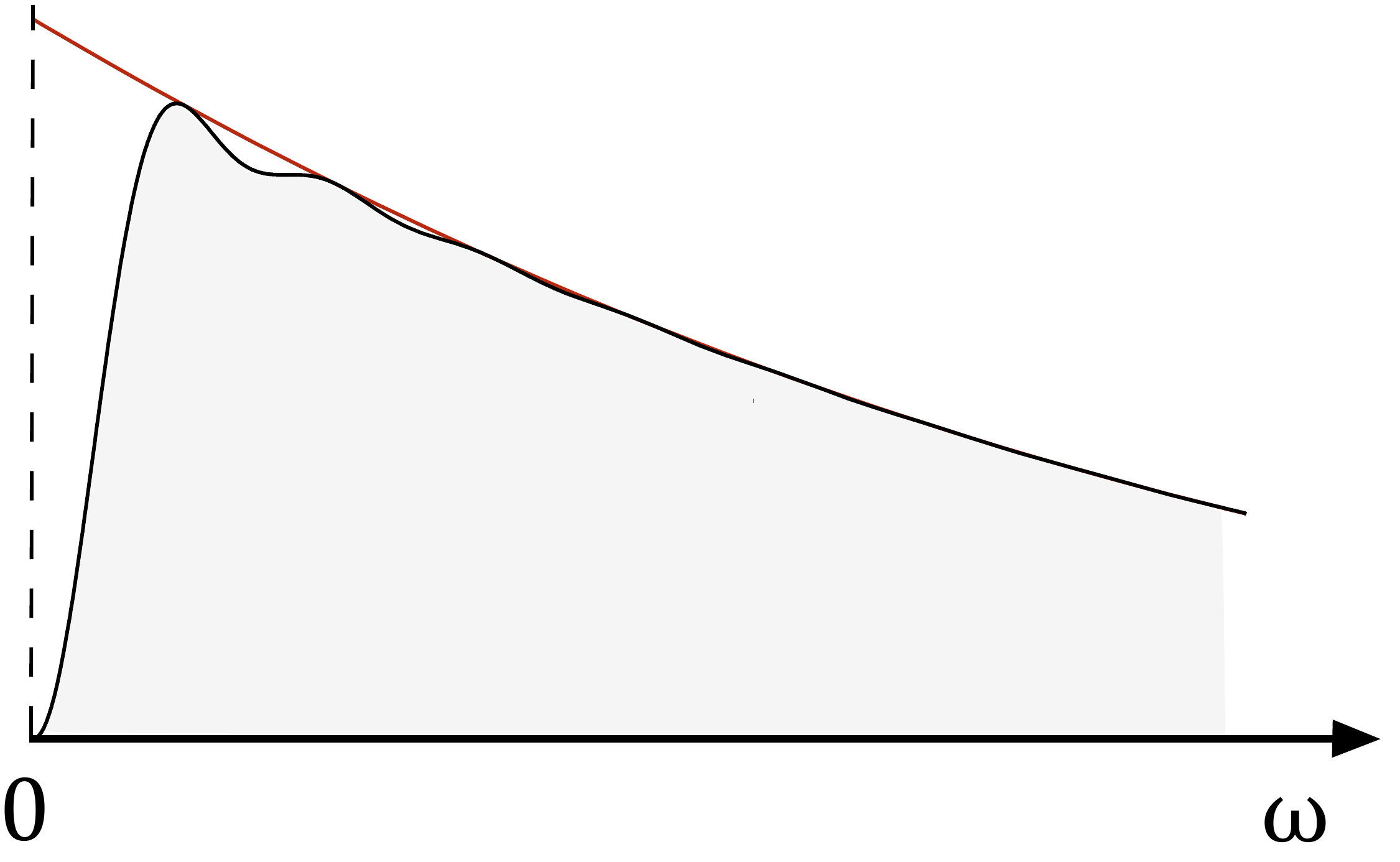}}\quad\label{234}
\end{equation}
The red curve denotes the semiclassical Planckian black body law with initial linear decay. At extremely low energies with $\omega$ of order $e^{-S_0}$ we explicitly see the depletion of the Hawking-Unruh spectrum due to level repulsion in quantum gravity.
\\~\\
To get a better grip on the physics we are probing we can briefly consider the bulk gravitational dual to one single quantum chaotic system with Hamiltonian $H$. In JT gravity, we can effectively achieve this by considering a new definition of JT gravity for which the path integral itself includes a bunch of additional branes \cite{sss2,paper6,maxfieldmarolf,wophilbert}. The data of these branes are one-to-one with the spectral data of the dual quantum mechanical system. For this setup in particular, we want to consider a version of JT gravity that includes branes which fix the eigenvalues $\lambda_1\dots \lambda_L$ of the Hamiltonians \cite{paper6}. Essentially we replace in \eqref{233} and \eqref{49}
\begin{equation}
    \rho(E,M)=\sum_{i=1}^L \delta(E-\lambda_i)\sum_{j=1}^L \delta(M-\lambda_j).\label{239}
\end{equation}
We imagine probing in a state of the coupled matter and gravity system with energy $M$ identical to one of the eigenvalues $\lambda$. We find via \eqref{51} and \eqref{239}
\begin{equation}
   R(\omega)= 2\,\frac{\sin^2 \omega z}{\omega^2} \, \sum_{i=1}^L \delta(\omega-\lambda+\lambda_i)\, \rvert \mo_{\lambda \lambda_i}^1\rvert^2.
\end{equation}
In the probe approximation $\lambda\gg 1$ and $\omega\ll \lambda$ this approximates to\footnote{Here $\rho(\lambda_i)$ is to be understood as a function that varies only on scales much larger than $e^{-S_0}$.}
\begin{equation}
R(\omega)_{\lambda} = 2\,\frac{\sin^2  \omega z}{\omega^2}\, \frac{\omega}{e^{\beta \omega}-1}\sum_{i=1}^L \frac{1}{\rho(\lambda_i)}\delta(\omega-\lambda+\lambda_i) .\label{55}
\end{equation}
This makes explicit formula \eqref{difference} which states that the energy levels $\omega$ of the Unruh-Hawking modes must match to energy differences of the quantum chaotic black hole system: the gravitationally dressed raising and lowering operators generate level transitions within the quantum gravitational system
\begin{equation}
    \raisebox{-15mm}{\includegraphics[width=48mm]{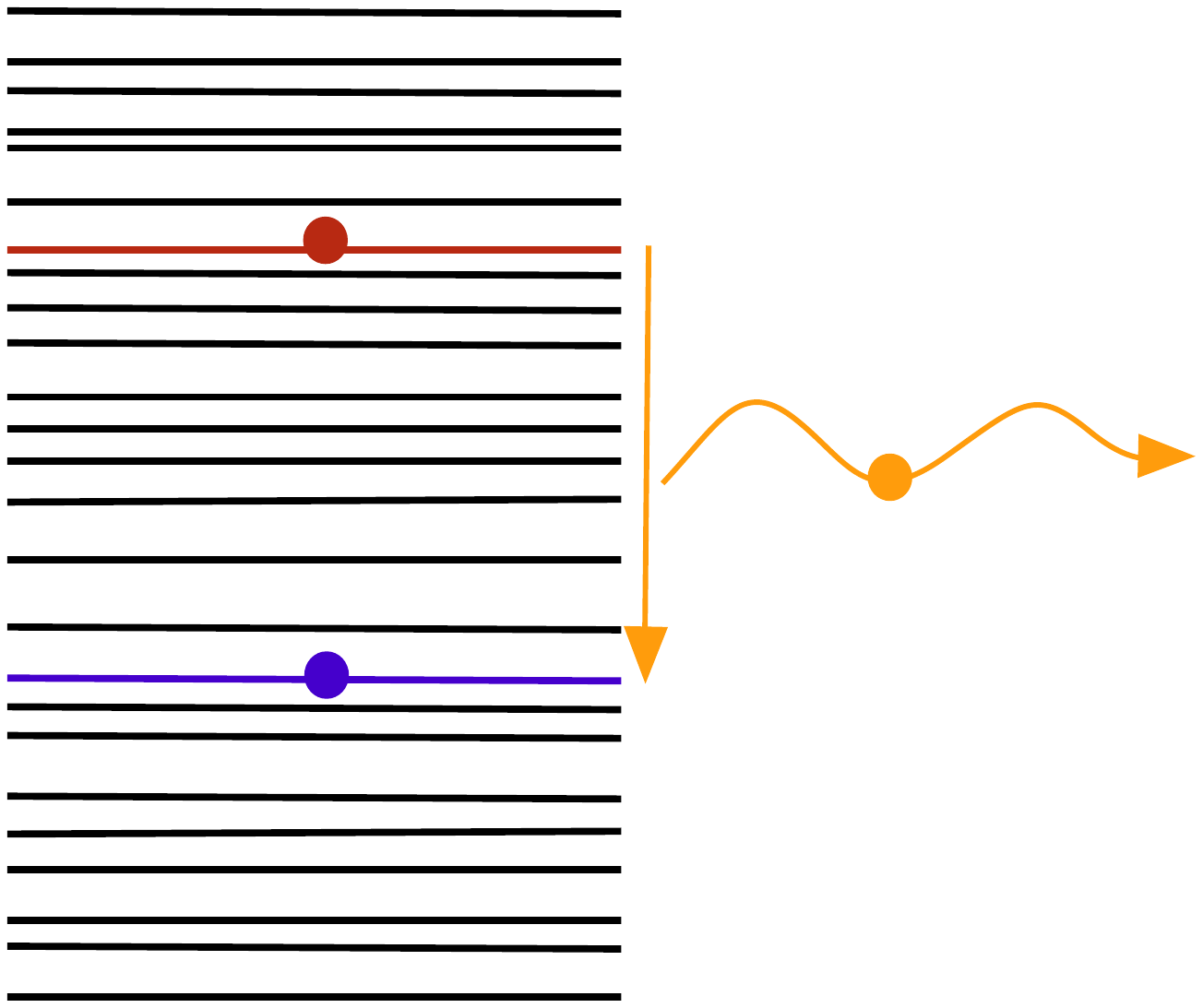}}\quad\label{242}
\end{equation}
We remove (orange) an energy difference $\omega_{ij}$ from an initial state (red) with energy $\lambda_i$ and are left with a final state (blue) with energy $\lambda_j$. 
\\~\\
The result \eqref{55} carries an interesting interpretation. It is commonly assumed that the detailed information about the microstate of a gravitational system is somehow located behind the horizon. In \eqref{55} however we have an explicit experiment which can be performed by an observer hovering outside of the semiclassical horizon. From the resulting set of delta spikes \eqref{55} as observed by his detector, he is in principle able to determine the energy spectrum of all the gravitational microstates. Here we take the details of one gravitational microstate to mean knowledge of all the levels $\lambda_1\dots \lambda_L$.

\subsection{Fermionic matter}
The result \eqref{expectationunruh} is expected to hold quite universally. We may provide evidence for this by testing it in more general situations. Therefore in this section we consider an Unruh-DeWitt detector coupled to massless fermionic bulk matter in JT gravity. In the next section \ref{s:proper} we consider other detector couplings.

We note that massless bulk fermions coupled to JT gravity are expected to play quite an important role in the elusive 2d gravity theory dual to the SYK model. Furthermore a large number of such massless bulk fermionic fields drastically enhance the evaporation rate for 4d magnetically charged black holes (which have a near horizon description as JT gravity) \cite{Maldacena:2018gjk,Maldacena:2020skw}.
\\~\\
We will consider as detector interaction the simplest possible coupling of a bosonic detector $\tilde{\mu}(t)$ to a bulk Dirac fermion $\psi(u,v)$ \cite{Takagi:1986tf}
\begin{align}
H_\text{int} &= g \, \tilde{\mu}(t) \, (\bar{\psi} \psi)(u(t),v(t)).\label{ferco}
\end{align}
See \cite{Hummer:2015xaa} for a comprehensive study of different Unruh-DeWitt detector types and their effects and \cite{Gray:2018ifq,Louko:2016ptn} for recent studies. The normal ordering defined by $(\bar{\psi} \psi) = \bar{\psi}\psi - \left\langle \bar{\psi}\psi\right\rangle_{\CFT}$ is required in order to sensibly define this detector coupling. The detector variable is chosen as
\begin{equation}
\label{rel}
\tilde{\mu}(t) = \frac{L_{\text{det}}}{\Omega(u(t),v(t))} \,\mu(t) \, .
\end{equation}
We recognize the same detector variable $\mu(t)$ from the bosonic detector \eqref{26}. Furthermore we introduced the Weyl factor $\Omega(u(t),v(t))$ of the metric
\begin{equation}
    ds^2=-\frac{1}{\Omega(u,v)^{2}}\,du\, dv = -\frac{F'(u)F'(v)}{(F(u)-F(v))^2} du\, dv \, .
\end{equation}
The relation \eqref{rel} can be motivated by dimensional analysis. The coupling $\mu(t)$ in the bosonic detector has units of inverse length whereas $\tilde{\mu}(t)$ is dimensionless. The Weyl factor $\Omega(u(t),v(t))$ transforms between the detector length scale $L_\text{det}$ and the local length scale.
We will suppress the detector length scale $L_{\text{det}}$. It is however not difficult to consider a detector with coupling that does not include this Weyl factor. We comment on this in section \ref{s:proper}.

Notice that because of the coupling to a composite operator in \eqref{ferco}, we are not probing for a fermionic emission spectrum, but instead are measuring the emission of a fermion-antifermion pair. This jives with the interpretation that we are probing with the Unruh-DeWitt detector the probability of level transitions within an underlying bosonic black hole system, for which level transitions are only possible upon emission of bosonic quanta. 
\\~\\
Let us develop this example in more detail. The 2d massless curved spacetime Dirac equation has the following Weyl rescaling property: if $\psi$ is a solution in the metric $g_{\mu\nu}$, then $\Omega^{1/2}\psi$ is a solution in the metric $\Omega^{-2}g_{\mu\nu}$. We can use this to write the mode expansion of the Dirac field in AdS$_2$ in \Poincare coordinates ($U,V$) as:\footnote{We use the Dirac algebra convention where $\gamma_0= i \sigma_1$ and $\gamma_1 = \sigma_3$, in terms of the Pauli $\sigma$-matrices.}
\begin{align}
\label{modefermi}
\frac{\psi(U,V)}{(U-V)^\frac{1}{2}}&=\psi(U)+\psi(V)\\&= \,\,\frac{1}{\sqrt{4\pi}}\,\left(\begin{array}{c} 1 \\ i \end{array}\right)\,\sum_{\omega>0} \left(e^{- i \omega U} a_{\omega} +e^{ i \omega U} b_{\omega}^{\dagger}\right) +\,\frac{1}{\sqrt{4\pi}}\,\left(\begin{array}{c} i \\ 1 \end{array}\right)\,\sum_{\omega>0} \left(e^{- i \omega V} a_{\omega} +e^{ i \omega V} b_{\omega}^{\dagger}\right) \, .\nonumber
\end{align}
We have implicitly adopted Dirichlet boundary conditions here.\footnote{Variation of the massless Dirac action gives a boundary condition:
\begin{equation}
\left. \bar{\psi} \gamma^1 \psi \right\rvert_\partial = 0 \, .
\end{equation}
This equation holds also in curved spacetime as one checks that all Weyl scaling factors end up cancelling.
In terms of spinor components $\psi_\alpha, \, \alpha=1,2$, this becomes:
\begin{equation}
\psi_1^*\, \psi_2\rvert_\partial = \psi_1 \,\psi_2^*\rvert_\partial \, .
\end{equation}
The Dirichlet boundary discussed above corresponds to:
\begin{equation}
    \psi_1\rvert_\partial =\psi_2\rvert_\partial \, .
\end{equation}
This results eventually in the $4\sin^2\,\omega z$ greybody interference factor in \eqref{260}. Other possibilities include for example setting $\psi_1=0$ or $\psi_2=0$. This results instead in a $4 \cos^2\, \omega z$ interference factor. See also appendix \ref{app:obc}.} 
Using this mode expansion, 
it is easy to obtain the Wightman bulk two point function in the \Poincare vacuum:
\begin{equation}
    \frac{\average{\psi_\alpha(U_1,V_1)\bar{\psi}_\beta(U_2,V_2)}_{\CFT}}{ (U_1-V_1)^\frac{1}{2}\,(U_2-V_2)^\frac{1}{2}} = S_{\alpha\beta}(U_1,V_1,U_2,V_2) \, ,
\end{equation}
where
\begin{equation}
    S(U_1,V_1,U_2,V_2)=\frac{1}{2}\left[ \begin{array}{cc}
\frac{1}{U_1-U_2} - \frac{1}{U_1-V_2} - \frac{1}{V_1-V_2} + \frac{1}{V_1-U_2} & \frac{i}{U_1-U_2} + \frac{i}{U_1-V_2} + \frac{i}{V_1-V_2} + \frac{i}{V_1-U_2} \\
\frac{i}{U_1-U_2} - \frac{i}{U_1-V_2} + \frac{i}{V_1-V_2} - \frac{i}{V_1-U_2} & -\frac{1}{U_1-U_2} - \frac{1}{U_1-V_2} + \frac{1}{V_1-V_2} + \frac{1}{V_1-U_2}
\end{array}\right] \nonumber \, .
\end{equation}
In terms of the detector we are led to compute the bulk two-point function of fermion-antifermion pairs. By taking Wick contractions, one finds 
\begin{align}
\label{contr}
\frac{\left\langle (\bar{\psi} \psi) (U_1,V_1) (\bar{\psi} \psi) (U_2,V_2)\right\rangle_{\CFT}}{(U_1-V_1)\,(U_2-V_2)}&= \sum_{\alpha\,\beta}\contraction{}{\overline{\psi}_\alpha}{\psi_\alpha \overline{\psi}_\beta}{\psi_\beta}
\contraction{\overline{\psi}_\alpha}{\psi_\alpha}{}{\overline{\psi}_\beta} \overline{\psi}_\alpha \psi_\alpha \overline{\psi}_\beta \psi_\beta = - \Tr S^2(U_1,V_1,U_2,V_2)\\&= -\frac{1}{(U_1-V_2)^2} -  \frac{1}{(V_1-U_2)^2} +  \frac{2}{(U_1-U_2)(V_1-V_2)}\nonumber \,.
\end{align}
Notice that by definition of normal-ordering, there are no contractions to be considered within each composite operator $(\bar{\psi}\psi)(U,V)$.
\\~\\
As in \eqref{215} for the bosonic case, we can couple this observable to the Schwarzian by applying a coordinate reparameterization and then computing the path integral. Before doing so, let us note that the Weyl factor in the detector couplings \eqref{ferco} cancels with the Weyl factors on the left hand side of \eqref{contr} in the detector transition rate \eqref{29}. So from hereon let us drop all such factors.

In reparameterized bulk metrics \eqref{189} and ignoring the Weyl factors, the fermion pair two-point function \eqref{contr} becomes:
\begin{align}
\label{contr2}
\nonumber &\average{(\bar{\Psi} \Psi) [f\vertrule u_1,v_1] (\bar{\Psi} \Psi) [f\vertrule u_2,v_2]}_{\CFT} \\&= -\frac{F'(u_1)F'(v_2)}{(F(u_1)-F(v_2))^2} -  \frac{F'(v_1)F'(u_2)}{(F(v_1)-F(u_2))^2} + 2\, \frac{F'(u_1)F'(u_2)}{(F(u_1)-F(u_2)}\,\frac{F'(v_1)F'(v_2)}{(F(v_1)-F(v_2)}\,,
\end{align}
in terms of the gravitationally dressed field $\Psi$.
The conformal scaling factors $F'$ are explained because the holomorphic and antiholomorphic components $\psi(u)$ and $\psi(v)$ in \eqref{modefermi} are $\ell=1/2$ conformal primaries. Following \eqref{29} and normalizing by the operator intrinsic prefactor on the first line of \eqref{29}, one now computes the response rate as:
\begin{align}
    R(\omega)=&\lim_{\T\to \infty}\frac{1}{\T}\int_{-\T}^{+\T}dt_1\int_{-\T}^{+\T}dt_2\,e^{-i\omega(t_1-t_2)}\nonumber\\&\qquad\qquad\qquad\qquad\bra{\mM}(\bar{\Psi} \Psi) [f \vertrule u_1(t_1),v_1(t_2)] (\bar{\Psi} \Psi) [f \vertrule u_2(t_1),v_2(t_2)]\ket{\mM}\, .\label{254}
\end{align}
As in the bosonic case we will consider a detector following a trajectory at fixed $z$. In implementing the Schwarzian path integral of \eqref{contr2} there are a few subtleties related with operator ordering ambiguities which we must first address.
\begin{enumerate}
    \item All three terms in \eqref{contr2} correspond to the product of two Schwarzian $\ell=1/2$ bilocals. 
		In our case the quantum mechanical operator ordering is fixed by the nested Wick contractions in \eqref{contr}, leading to three nested Schwarzian four-point functions.
    \item 
		The first two terms of \eqref{contr2} correspond to two Schwarzian bilocals with the same start and end points. The Schwarzian correlator can be simplified using the following property:
    \begin{equation}
        \quad\raisebox{-10mm}{\includegraphics[width=45mm]{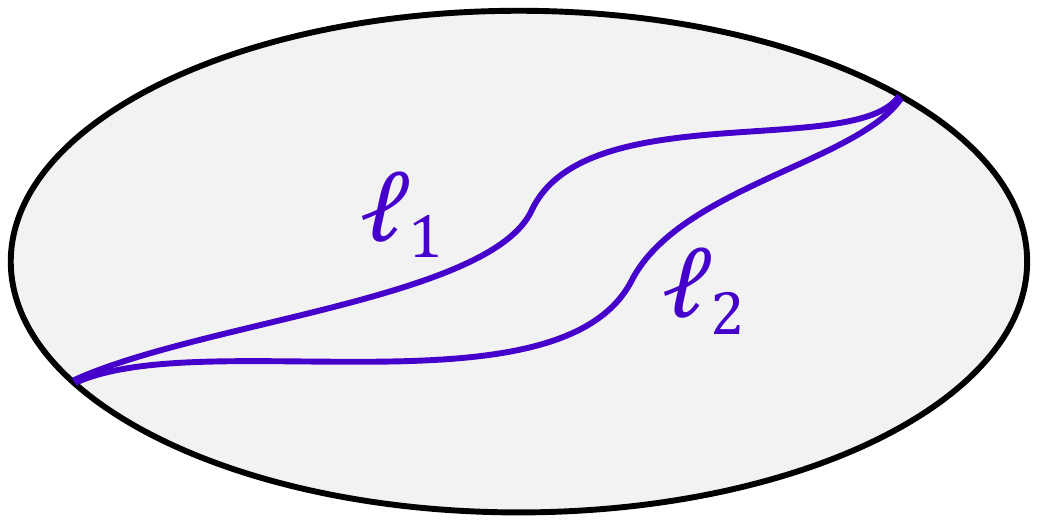}}\quad=\quad\raisebox{-10mm}{\includegraphics[width=45mm]{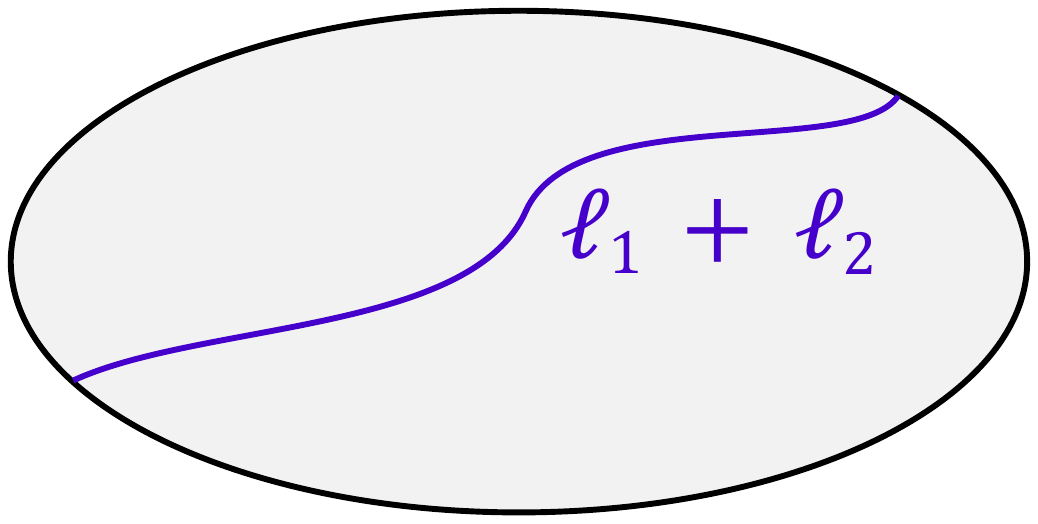}}\quad.\label{256}
    \end{equation}
    Both these JT gravity amplitudes can be immediately evaluated using the result of \cite{schwarzian,Mertens:2018fds,paper3,Blommaert:2018iqz}. 
		Labeling the energies as $E_1$, $E$ and $E_2$ from left to right of the diagram \eqref{256}, we have the result:\footnote{This follows from a generalization of the Barnes identities and is identical to the orthonormality relation of the zeroth Wilson polynomial \cite{groenevelt}. In particular the identity is:
    \begin{align}
        \frac{1}{4\pi i}\int_{-i\infty}^{+i\infty}ds \frac{\Gamma(a \pm s) \Gamma(b \pm s) \Gamma( c \pm s) \Gamma (d\pm s)}{\Gamma(\pm 2s)} = \frac{\Gamma(a+b)\Gamma(a+c)\Gamma(a+d)\Gamma(b+c)\Gamma(b+d)\Gamma(c+d)}{\Gamma(a+b+c+d)}.  
				 \end{align}
				This identity holds when $\Re a,b,c,d >0$. We checked it numerically as well.
       }
    \begin{align}
        \int_0^\infty d E\, \rho_0(E)\, \rvert\mo^{\ell_1}_{E_1 E}\rvert^2\,\rvert\mo^{\ell_2}_{E E_1}\rvert^2 = \rvert\mo^{\ell_1+\ell_2}_{E_1 E_2}\rvert^2\, .\label{257}
    \end{align}
		This means we can use the classical identity
		\begin{equation}
				\left(\frac{F'_1 F'_2}{(F_1-F_2)^2} \right)^{\ell_1 } \left(\frac{F'_1 F'_2}{(F_1-F_2)^2} \right)^{\ell_2 } = \left(\frac{F'_1 F'_2}{(F_1-F_2)^2} \right)^{\ell_1+\ell_2 } \, ,
    \end{equation}
		also at the quantum level. We also have the more general identity:
    \begin{equation}
        \quad\raisebox{-10mm}{\includegraphics[width=45mm]{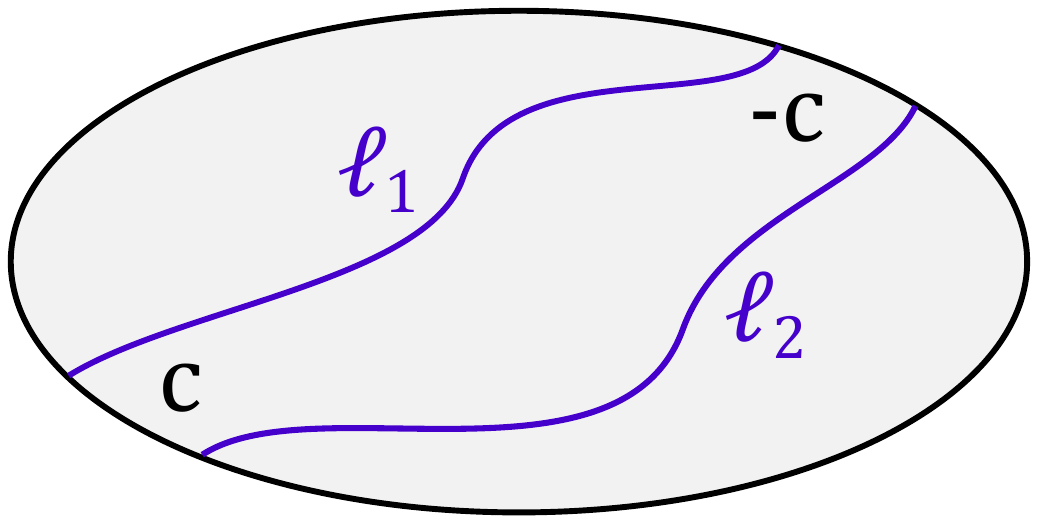}}\quad=\quad\raisebox{-10mm}{\includegraphics[width=45mm]{l1plus.pdf}}\quad.\label{259}
    \end{equation}
		This property also reduces the Schwarzian path integral of the last term in \eqref{contr2} (for which $c=2i z$) to a single bilocal computation.
\end{enumerate}
Proceeding by the calculation of \eqref{254} we find that this boils down to computing three Schwarzian diagrams of the same type. 
Their prefactors combine as an interference factor as $ - \frac{1}{2}e^{i\omega 2z} -  \frac{1}{2}e^{-i\omega 2z} + 1 = 2 \sin^2 \omega z$, leading to the result
\begin{equation}
R(\omega) = 2\,\sin^2 \omega z\, \rho_0(M-\omega)\, \rvert\mo_{M M-\omega}^1 \rvert^2\,.\label{260}
\end{equation}
Finally, we should include Euclidean wormholes to this computation. The calculation is essentially identical to the one which led to \eqref{51}. We note that \eqref{256} does not technically hold when including such Euclidean wormholes because we could have Euclidean wormholes connecting the middle region with the two other regions. 
Such corrections are important in the intermediate $E$ integral whenever $E \approx M$ and or $E\approx M-\omega$ where the contribution to the $E$ integral is pushed to zero by quadratic level repulsion. This region has energy width $\sim e^{-S_0}$ and order $1$ height. Doing the $E$ integral results in an order $e^{-S_0}$ correction to the response rate. This is negligible. In other words we may safely still use \eqref{256} even when including Euclidean wormhole contributions to both diagrams.

The final result is hence essentially identical to the bosonic answer \eqref{51} and provides a second example of our generic expectation \eqref{expectationunruh} in the probe approximation $M\gg 1$ and $\omega \ll M$. The difference sits only in the interference factor which differs by an overall $\omega^2$. This comes purely from dimensional reasons since there is also a length scale $L_{\text{det}}$ in the coupling \eqref{ferco}. The relative prefactor between \eqref{260} and \eqref{response} is the dimensionless combination $L_{\text{det}}^2 \, \omega^2$.

\subsection{More general detector couplings}
\label{s:proper}
The detector couplings $\mu(t)$ and $\tilde{\mu}(t)$ we have defined previously, transform as scalar densities under coordinate transformations, and correspond to time measurements on the boundary clock. It is straightforward to consider coupling to the proper bulk time of the worldline, by changing the interaction terms into
\begin{align}
\label{propcou}
   S_\text{int} &= g\int d\tau_p \, \mu(\tau_p)\, \phi(u(\tau_p),v(\tau_p)) = g\int dt \, \Omega(u(t),v(t))^{-1} \, \mu(t)\, \phi(u(t),v(t)) \, ,  \\
   S_\text{int} &= g\int d\tau_p \, \tilde{\mu}(\tau_p) \, (\bar{\psi} \psi)(u(\tau_p),v(\tau_p)) = g  \int dt \,  \Omega(u(t),v(t))^{-1} \, \tilde{\mu}(t) \, (\bar{\psi} \psi)(u(t),v(t)) \, . \nonumber 
\end{align}
Here $dt \, \Omega(u(t),v(t))^{-1} = d\tau_p$ is the proper time along the bulk worldline. At the quantum gravity level, the above coupling is replaced by a Hermitian coupling including the dressed fields
\begin{align}
   S_\text{int} = \frac{g}{2}\int dt \,  \mu(t) \Bigl( \Omega(u(t),v(t))^{-1} \,\Phi[f\vertrule u(t),v(t))] + \Phi[f\vertrule u(t),v(t))]\Omega(u(t),v(t))^{-1} \Bigr)  \nonumber \, .
\end{align}
Notably also $\Omega^{-1}$ depends on the gravity variable $f$. An analogous formula holds for the fermion detector.  The explicit Weyl factors $\Omega^{-1}$ in this expression yield additional Schwarzian bilocal lines with length $2z$ between their endpoints. Diagrammatically we have:
\begin{equation}
	\label{weyl}
       \frac{1}{2} \quad\raisebox{-10mm}{\includegraphics[width=45mm]{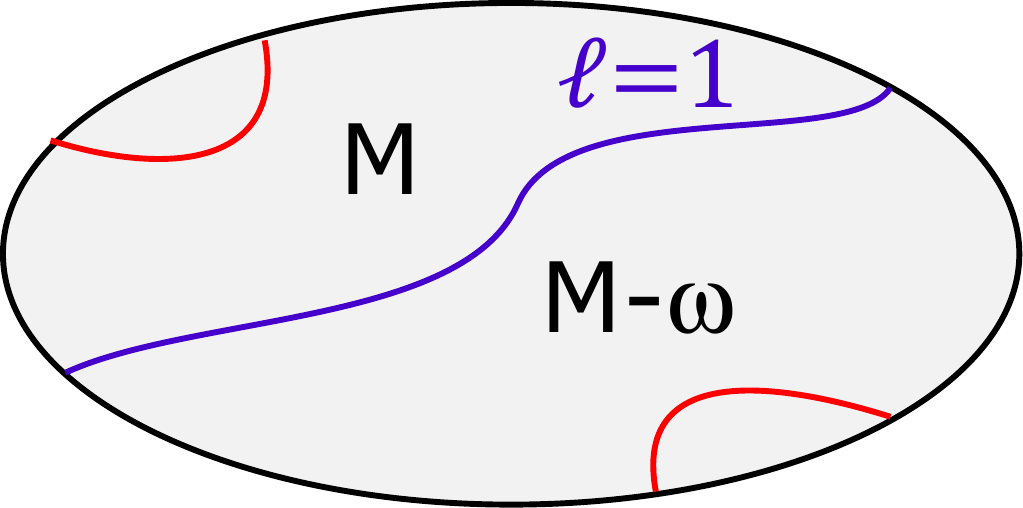}}\quad+ \frac{1}{2}\quad\raisebox{-10mm}{\includegraphics[width=45mm]{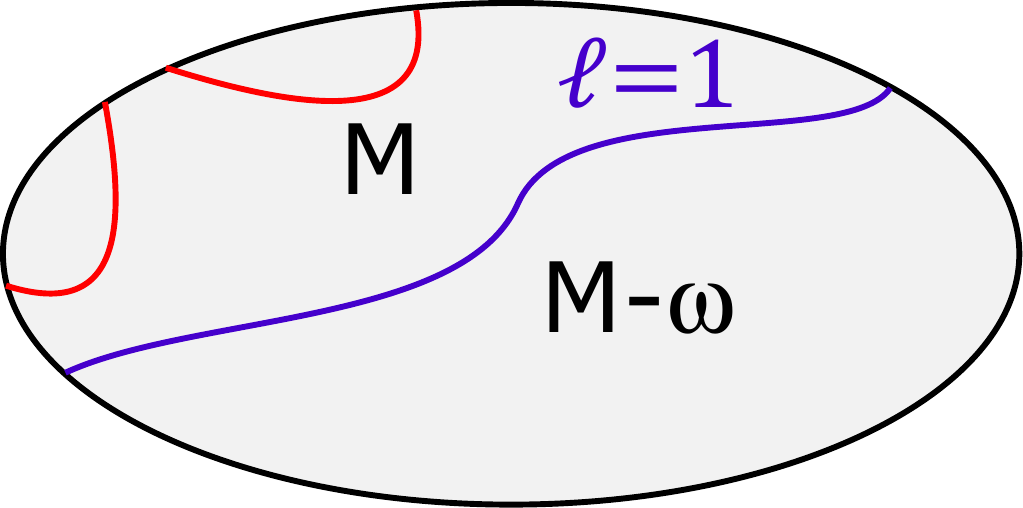}}\quad.
\end{equation}
The bilocal lines coming from the explicit Weyl factors are represented by the red lines. The important point is that since this additional structure only depends on $z$, it does not participate in the Fourier transform over $t$, and factorizes from the amplitudes. Furthermore Euclidean wormholes connecting to these new regions in \eqref{weyl} only give subdominant contributions because we are integrating over the corresponding energy labels of these regions. Hence the only effect of including these Weyl factors in \eqref{26} is that \eqref{response} and \eqref{51} receive an additional overall prefactor which depends on $z$ but which is crucially independent of $\omega$. By consequence one still finds the same physics as in for example \eqref{234}. The fermionic coupling requires exactly the same treatment with Weyl factors on either side of the $\ell=1/2$ pair of lines in \eqref{256}.

We note to conclude that one could consider more generic detector couplings, of higher order in the fields, such as:
\begin{equation}
H_\text{int} = g\,\mu(t)\,f(\phi(u(t),v(t))\quad,\quad f(\phi(u(t),v(t)) = \sum_{n} c_n (\phi^n)(u(t),v(t)) \, .
\end{equation}
The precise Schwarzian and Euclidean wormhole computations are more involved. The resulting detector response would be obtained as the same Taylor series expansion:
\begin{equation}
R_g(\omega)=\sum_n c_n R_n(\omega) \, .
\end{equation}
Intuitively, we expect to see level repulsion in any such response rate given its universal role in random matrix correlation functions.

\section{Energy density in the Unruh heat bath}
\label{s:hb}
In this section we investigate the spectral energy density in the Unruh heat bath. Semi-classically this is given by $\bra{\mM}\omega\,a_\omega^\dagger a_\omega \ket{\mM}$, so we might expect the gravitational corrections to be accounted for by considering instead $\bra{\mM}\omega\,A_\omega^\dagger[f] A_\omega[f] \ket{\mM}$. However, this is not true. This traces back to operator ordering ambiguities when promoting quantum field observables in a fixed metric \eqref{189} to operators in a theory of matter-coupled quantum gravity.
\\~\\
Let us first focus on coupling to the Schwarzian reparametrization. The local energy density in the Unruh heat bath can be computed independently using the coincident limit of the bulk two-point function (regularized via point splitting) \cite{spradlin}. The result is:
\begin{equation}
\label{Unruhflux}
\left\langle :T_{uu}[f \vertrule u]:\right\rangle_{\CFT} = -\frac{1}{24\pi}\text{Sch}\left(\tanh\frac{\pi}{\beta}f,u \right), \quad \left\langle :T_{vv}[f \vertrule v]:\right\rangle_{\CFT} = -\frac{1}{24\pi}\text{Sch}\left(\tanh\frac{\pi}{\beta}f,v\right) \, .
\end{equation}
Doing the gravitational path integral one finds \cite{Mertens:2019bvy}
\begin{align}
\label{Unruhflux2}
\bra{\mM}:T_{uu}[f \vertrule u]:\ket{\mM} &= \bra{\mM}:T_{vv}[f \vertrule v]:\ket{\mM}  =\frac{M}{12 \pi}\, .
\end{align}
The total energy in the heat bath is then\footnote{We have included the contribution from the $v$-lightcone component by using the doubling trick.}
\begin{equation}
E_{\text{bath}} = \int_{-\infty}^{+\infty} du\, \bra{\mM}:T_{uu}[f \vertrule u]:\ket{\mM}=\frac{V M}{12\pi}\, .\label{33}
\end{equation}
The bath energy $E_{\text{bath}}$ is defined operationally by summing the bath energy densities $T_{uu}$ across an entire spatial slice. However, this is not the total matter energy that backreacts on the geometry: including the vacuum energy gives a net zero result. In flat spacetime this famously leads to the statement that the energy in the Unruh heat bath does not deform the flat geometry one started with. 
\\~\\
Our goal here is to find a spectral quantity that reproduces this bath energy $E_{\text{bath}}$, and the first candidate is $\bra{\mM}\omega\,A_\omega[f]^\dagger A_\omega[f] \ket{\mM}$. However, one finds very explicitly from \eqref{num1} that:
\begin{equation}
\int_{0}^{+\infty}d\omega \bra{\mM}\omega\,A_\omega^\dagger[f] A_\omega[f] \ket{\mM}\, \neq \, E_{\text{bath}}\, .\label{34}
\end{equation}
The potential confusion is closely related to the fact that the operators $A_\omega$ constructed in \eqref{creabn} are not quite the same as raising and lowering operators acting on the matter Hilbert space (nor where they constructed to be). For example, we may compute:
\begin{align}
\bra{\mM}[A_{\omega_1},A^\dagger_{\omega_2}]\ket{\mM} =  \frac{\delta(\omega_1-\omega_2)}{\omega_1}\,&\rho_0(M+\omega_1)\,\rvert\mo_{M \, M+\omega_1}^1\rvert^2\nonumber \\
&-\frac{\delta(\omega_1-\omega_2)}{\omega_1}\,\rho_0(M-\omega_1)\,\rvert \mo_{M\,  M-\omega_1}^1\rvert^2\, .\label{35}
\end{align}
In the probe approximation $M\gg 1$ and $\omega\ll M$ this does reduce to $\delta(\omega_1-\omega_2)$. However for more generic values of $M$ and $\omega$, it does not.

We will below define the spectral energy density $\omega N_\omega[f]$ that does satisfy this relation:
\begin{equation}
\int_{0}^{+\infty}d\omega \bra{\mM}\omega N_\omega[f]\ket{\mM} = E_{\text{bath}} \, ,
\end{equation}
and find that it corresponds to a symmetrized version of $\omega A^{\dagger}_\omega A_\omega$.

\subsection{Symmetric number operator}
The goal here is to introduce a symmetric number operator $N_\omega[f]$, which does compute the bath spectral energy density. Consider first semi-classical matter. Following \eqref{creaan} one defines a number operator
\begin{equation}
\average{a^\dagger_\omega a_\omega}_{\CFT} =  \frac{1}{\pi\omega}\int_{-\infty}^{+\infty} du_1\int_{-\infty}^{+\infty}du_2\, e^{-i\omega(u_1-u_2)}\, \average{\partial\phi(u_1) \partial\phi(u_2)}_{\CFT}\, .\label{38}
\end{equation}
This features the Wightman bulk two-point function. Using the canonical commutation relation
\begin{equation}
 a^\dagger_\omega a_\omega =\frac{1}{2}a^\dagger_\omega a_\omega+\frac{1}{2} a_\omega a^\dagger_\omega -\frac{1}{2}\delta(0)\, ,\label{39}
\end{equation}
we can write this in a more symmetric way as:
\begin{align}
\label{symoper}
\average{a^\dagger_\omega a_\omega}_{\CFT} &=  \frac{1}{\pi\omega}\int_{-\infty}^{+\infty} du_1\int_{-\infty}^{+\infty}du_2\, e^{-i\omega(u_1-u_2)}\, \average{\partial\phi(u_1) \partial\phi(u_2)}_{\CFT}\nonumber \\&\qquad+ \frac{1}{\pi\omega}\int_{-\infty}^{+\infty} du_1\int_{-\infty}^{+\infty}du_2\, e^{-i\omega(u_1-u_2)}\, \average{\partial\phi(u_2) \partial\phi(u_1)}_{\CFT} - \frac{1}{2} \delta(0)\,.
\end{align}
Semiclassically, this manipulation is completely harmless. The point is that \eqref{38} and \eqref{symoper} turn out not to be equivalent when promoting the Wightman two-point functions to operators in quantum gravity. This has to do with operator orderings.

The two Wightman bulk two-point functions are not precisely identical. Including an appropriate $i\varepsilon$ regulator we have
\begin{align}
\label{wight}
    \average{\partial\phi(u_1) \partial\phi(u_2)}_{\CFT}&=-\frac{1}{4\pi}\,\frac{1}{\frac{\beta^2}{\pi^2}\sinh^2 \frac{\pi}{\beta}(u_1-u_2+i\epsilon)}, \nonumber\\ \average{\partial\phi(u_2) \partial\phi(u_1)}_{\CFT}&=-\frac{1}{4\pi}\,\frac{1}{\frac{\beta^2}{\pi^2}\sinh^2\frac{\pi}{\beta}(u_1-u_2-i\epsilon)}\,.
\end{align}
In Schwarzian quantum gravity, we can then likewise define the symmetrized version of the dressed operators $A^{\dagger}_\omega A_\omega$ and $A_\omega A^{\dagger}_\omega$ as:
\begin{align}
 B^\dagger_\omega[f] B_\omega[f] &= \frac{1}{2}A^\dagger_\omega[f] A_\omega[f] + \frac{1}{2}A_\omega[f] A^\dagger_\omega[f] -\frac{1}{2}\delta(0)\, ,\\
 B_\omega[f] B^\dagger_\omega[f] &= \frac{1}{2}A^\dagger_\omega[f] A_\omega[f] + \frac{1}{2}A_\omega[f] A^\dagger_\omega[f] + \frac{1}{2}\delta(0)\, .
\end{align}
By construction these modes $B_\omega$ automatically satisfy the commutator relation:
\begin{equation}
[B_{\omega_1}[f],B_{\omega_2}^\dagger[f]]=\delta(\omega_1-\omega_2)\, .\label{313}
\end{equation}

Dressing the bulk Wightman two-point functions in \eqref{symoper} to include for Schwarzian gravitational interactions as in \eqref{218}, we write for $N_\omega[f] = B^\dagger_\omega B_\omega$ \cite{Mertens:2019bvy}:
\begin{align}
\label{planckp}
&\average{N_\omega[f]}_{\CFT} \\\nonumber &= -\frac{1}{8\pi^2\omega}\int_{-\infty}^{+\infty} d u_1 \int_{-\infty}^{+\infty} du _2\, e^{-i\omega (u_1-u_2)}\,\frac{f'(u_1)f'(u_2)}{\frac{\beta^2}{\pi^2}\sinh^2 \frac{\pi}{\beta}(f(u_1)-f(u_2)+i\varepsilon)} - \frac{1}{(u_1-u_2+i\varepsilon)^2}\\\nonumber &\quad -\frac{1}{8\pi^2\omega}\int_{-\infty}^{+\infty} d u_1 \int_{-\infty}^{+\infty} du _2\, e^{-i\omega (u_1-u_2)}\,\frac{f'(u_1)f'(u_2)}{\frac{\beta^2}{\pi^2}\sinh^2 \frac{\pi}{\beta}(f(u_1)-f(u_2)-i\varepsilon)} - \frac{1}{(u_1-u_2-i\varepsilon)^2}\,.
\end{align}
Here we have rewritten the delta-function in \eqref{symoper} as a term which explicitly subtracts the poles in the Wightman two-point functions.\footnote{This is directly found by using the integral representation of the step function. See also \cite{Fabbri:2004yy} for the semi-classical versions of these equations.} With this operator, we first compute the energy:
\begin{equation}
    \int_0^\infty d\omega\,\average{\omega N_\omega[f]}_{\CFT} \, .
\end{equation}
By swapping the integration variables $u_1$ and $u_2$ in the integrals of \eqref{planckp}, we see that the terms on the middle and last line are mapped into one another if we change the sign of $\omega$. Therefore, including a $1/2$ factor to enlarge the integration over $\omega$ along the entire real axis, we obtain a factor
\begin{equation}
    \int_{-\infty}^{+\infty} d\omega\,e^{-i\omega(u_1-u_2)}=2\pi \delta(u_1-u_2) \, .
\end{equation}
One then finds directly from Taylor expanding \eqref{planckp}\footnote{The validity of such a series expansion within Schwarzian correlators is not entirely straightforward. We refer to \cite{schwarzian,Mertens:2019tcm} for some comments and to \cite{toth} for a thorough analysis.}
\begin{align}
- \frac{1}{8\pi} \int_{-\infty}^{+\infty} du_1\int_{-\infty}^{+\infty} d u_2 \,\delta(u_1-u_2)\,&\frac{f'(u_1)f'(u_2)}{\frac{\beta^2}{\pi^2}\sinh^2 \frac{\pi}{\beta}(f_1-f_2-i\epsilon)} - \frac{1}{(u_{1}-u_{2}-i\epsilon)^2} + (\epsilon \to - \epsilon) \nonumber \\&\qquad\qquad=  - \frac{1}{24\pi} \int d y\, \text{Sch}\left(\tanh \frac{\pi}{\beta}f,y\right) \, .
\end{align} 
As classical functions this is true in any case. However, in the Schwarzian path integral this is only true when we work with the symmetrically dressed operators $B_\omega$. The reason is that the two Wightman two-point functions are not equal to one another after the Schwarzian path integral.\footnote{Unlike the semi-classical answer \eqref{wight}, the Schwarzian answers \eqref{41} and its complex conjugate differ by more than an infinitesimal term.} The above trick $\omega \to - \omega$ maps one into the other, but since we started with a symmetric combination this has no impact. This confirms that we may view the Schwarzian path integral of \eqref{planckp} as the spectral energy density $\bra{\mM}\omega N_\omega[f] \ket{\mM}$ in the Unruh heat bath. This agrees with the formulas presented in \cite{Mertens:2019bvy}.

\subsection{Analytical analysis}
Doing the Schwarzian path integrals in \eqref{planckp}, we find
\begin{align}
    \bra{\mM}\omega N_\omega[f]\ket{\mM} = \frac{V}{4\pi}\,\rho_0(M+\omega)\,\rvert \mo_{M M+\omega}^1\rvert^2+\frac{V}{4\pi}\,\rho_0(M-\omega)\,\rvert \mo_{M M-\omega}^1\rvert^2-\frac{\omega}{2}\,\delta(0) \, .\label{319}
\end{align}
To deal with the delta-function, we will choose to calibrate our measurement to the zero energy \Poincare state\footnote{Other options exist, and will be explored elsewhere.}
\begin{equation}
    \bra{0}\omega N_\omega[f]\ket{0}=\frac{V}{4\pi}\,\rho_0(\omega)\,\rvert \mo_{0\, 0+\omega}^1\rvert^2+\frac{V}{4\pi}\,\rho_0(-\omega)\,\rvert \mo_{0\,0-\omega}^1\rvert^2-\frac{\omega}{2}\,\delta(0)\, .\label{320}
\end{equation}
We will henceforth only discuss this relative spectral energy density
\begin{align}
    \bra{\mM}\omega N_\omega[f]\ket{\mM}=&\frac{V}{4\pi}\,\rho_0(M+\omega)\,\rvert \mo_{M M+\omega}^1\rvert^2-\frac{V}{4\pi}\,\rho_0(\omega)\,\rvert \mo_{0\, 0+\omega}^1\rvert^2\nonumber\\&+\frac{V}{4\pi}\,\rho_0(M-\omega)\,\rvert \mo_{M M-\omega}^1\rvert^2-\frac{V}{4\pi}\,\rho_0(-\omega)\,\rvert \mo_{0\,0-\omega}^1\rvert^2 \, .\label{321}
\end{align}
In the semi-classical probe approximation $M\gg 1$ and $\omega\ll M$ one recovers the classical answer for the spectral energy density
\begin{equation}
    \bra{\mM}\omega N_\omega[f]\ket{\mM}=\frac{V}{2\pi}\frac{\omega}{e^{\beta\omega}-1}\, .\label{322}
\end{equation}
The Schwarzian answer \eqref{321} which combines \eqref{319} and \eqref{320} is accurate for any $M\gg e^{-2S_0/3}$ and for any $\omega\gg e^{-S_0}$. This includes Planck sized black holes where $M\sim 1$. For such tiny black holes and with $\omega < M$ one finds slightly lower spectral energy density in \eqref{321} as compared to \eqref{322}. See figure \ref{Urandomz2}. On the other hand for $\omega>M$ the Schwarzian result \eqref{321} gives a slightly higher occupation as compared to the classical answer \eqref{322}.

Including Euclidean wormhole corrections to the Schwarzian correlations is done analogously as in \eqref{51}, and leads to:
\begin{align}
     \nonumber \bra{\mM}\omega N_\omega[f]\ket{\mM}=&\frac{V}{4\pi}\,\frac{\rho(M,M+\omega)}{\rho(M)}\,\rvert \mo_{M M+\omega}^1\rvert^2-\frac{V}{4\pi}\,\frac{\rho(0,0+\omega)}{\rho(0)}\,\rvert \mo_{0\,0+\omega}^1\rvert^2\nonumber\\&+\frac{V}{4\pi}\,\frac{\rho(M,M-\omega)}{\rho(M)}\,\rvert \mo_{M M-\omega}^1\rvert^2-\frac{V}{4\pi}\,\frac{\rho(0,0-\omega)}{\rho(0)}\,\rvert \mo_{0\,0-\omega}^1\rvert^2\, .\label{323}
\end{align}
The behavior of this function is in fact quite similar to \eqref{234}. In particular we see quadratic level repulsion for $\omega \sim e^{-S_0}$. One notable difference with \eqref{234} is the effect of the zero energy subtractions. As we review in appendix \ref{s:zeroref} these come which much slower wiggles as compared to the oscillations in \eqref{234}. The last term in \eqref{323} is in the forbidden region and is suppressed as $\sim e^{-S_0}$ in any case, making it negligible in practice. Supersymmetric JT gravity and the resulting random matrix completion from the Altland-Zirnbauer ensembles have a hard spectral edge at $\omega=0$ \cite{Stanford:2019vob} removing the forbidden region alltogether.

We have checked numerically for \eqref{321} that this indeed computes the spectral energy density \eqref{33}, by satisfying
\begin{equation}
\label{toten}
\int_0^{+\infty} d\omega \bra{\mM}\omega N_\omega[f]\ket{\mM} = \frac{V M}{12\pi} \,.
\end{equation}
Upon including higher topology in \eqref{323}, there are several tiny corrections. This is perfectly fine since neither \eqref{Unruhflux2} nor \eqref{33} is precise when including Euclidean wormhole corrections. We note that via \eqref{49} one finds that such corrections are at least suppressed by a factor $e^{-S_0}$. Therefore they can be neglected.\footnote{Actually, for all intents and purposes this suppression is exact. The sine kernel and contact term contributions in \eqref{49} cancel perfectly upon integration (by construction). Furthermore there are only quick wiggles but their integral gives a suppressed effect as well.}

For completeness, we note that the contact term contribution in \eqref{49} implies a further contribution to \eqref{323} of the form:
\begin{align}
\bra{\mM}\omega N_\omega[f]\ket{\mM} \, \supset \,  \delta(\omega)\,\frac{V}{2\pi}\, \bigl(\rvert \mo_{M M}^1\rvert^2-\,\rvert\mo_{0\,0}^1\rvert^2 \, \bigr) \,.
\end{align}
It is impossible to measure precisely zero energy so this contribution seems less interesting. On the other hand, there are several known examples of zero energy modes being important in a black hole context \cite{Donnelly:2014fua,Donnelly:2015hxa,Blommaert:2018rsf,Blommaert:2018oue}.

\subsection{Numerical analysis}
\begin{figure}[H]
\centering
\includegraphics[width=0.95\textwidth]{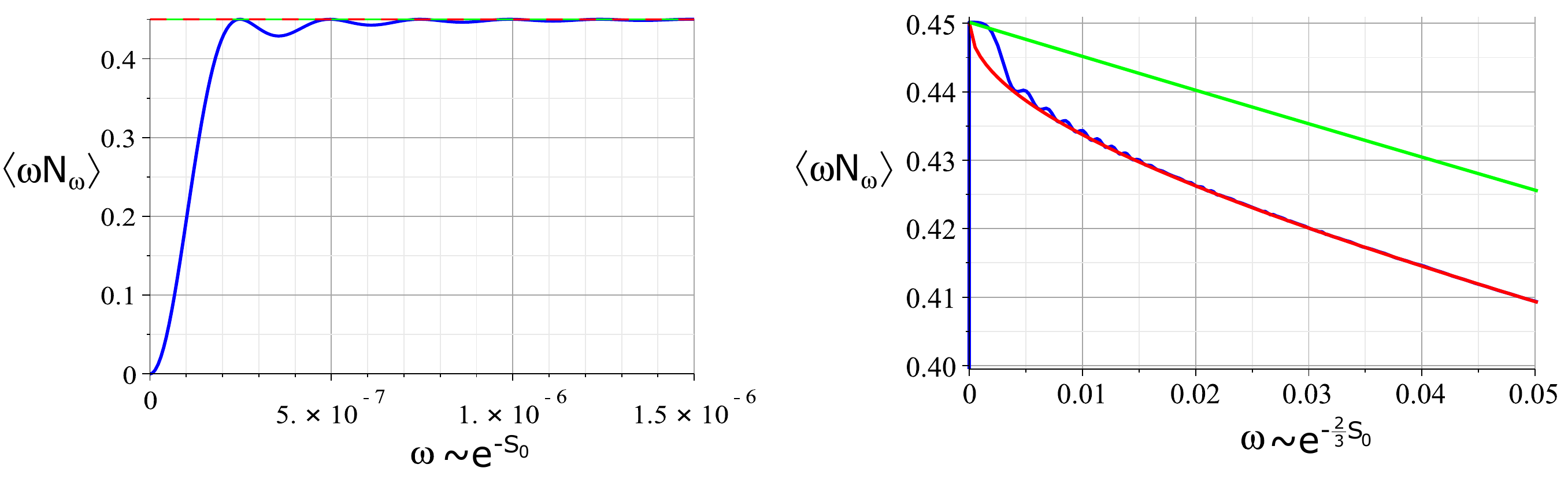}
\caption{(left) Zoom-in on the region with $\omega$ or order $e^{-S_0}$ where we see level repulsion and high-frequency wiggles in the exact (blue) result. (right) Zoom-in on the region $\omega$ of order $e^{-2S_0/3}$ where we see the slower Airy wiggles from the zero energy reference \eqref{323}. Furthermore one sees that the Schwarzian answer \eqref{321} (red) is lower than the semi-classical answer \eqref{322} (green).}
\label{Urandomz2}
\end{figure}

\begin{figure}[H]
\centering
\includegraphics[width=0.95\textwidth]{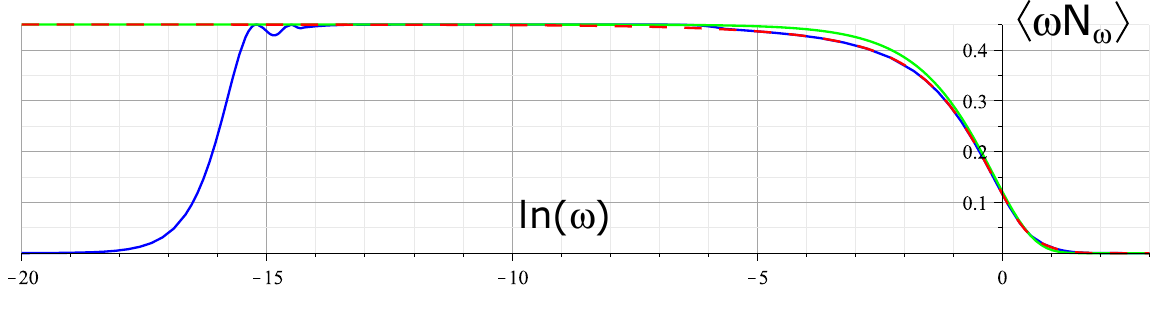}
\caption{The Schwarzian effects (right) and level repulsion effects (left) are simultaneously visible in this log plot of $\bra{\mM}\omega N_\omega[f]\ket{\mM}$. One may compare the exact answer (blue) with the Schwarzian answer (red) and the semiclassical answer (green).}
\label{Urandomfull}
\end{figure}

It is clarifying to explicitly plot the spectral energy density as computed in the three levels of improving approximation in \eqref{322}, \eqref{321} and \eqref{323}. We note that the zero-energy kernels in \eqref{323} are different from the universal answer \eqref{49}, which only holds far enough from the spectral edge. For the zero-energy kernels we may utilize instead results from the exactly solvable Airy model.\footnote{The kernel \eqref{49} is accurate as long as both $E, M\gg e^{-2S_0/3}$.} We present some details in appendix \ref{s:zeroref}.
\\~\\
The effects which we aim to see in the plots are
\begin{enumerate}
    \item The main effect of the Schwarzian corrections is that the Schwarzian answer (red) lies below the semiclassical answer (green) for small $\omega$ and above the semiclassical answer for large $\omega$.
    \item Secondly and most importantly, we can clearly see the signs of level repulsion. There is a depletion in the spectral density of the Unruh heat bath for $\omega\ll e^{-S_0}$ and furthermore we see high-frequency wiggles in the regime where $\omega$ is order $e^{-S_0}$.
    \item Finally, there are similar wiggles associated to the zero-energy subtraction in \eqref{323}. These will play a role when $\omega$ is order $e^{-2S_0/3}$. For $M\gg 1$ these Airy wiggles effectively become invisible as all contributions to \eqref{323} grow exponentially with $M$. 
\end{enumerate}
We will consider a parametric regime where all these effects are clearly visible. Therefore we take $M=2$ and $S_0=10$ in both figures \ref{Urandomz2} and \ref{Urandomfull}. 
\section{Concluding remarks}
\label{s:concl}
The main goal of this work was to advocate and provide evidence that Hawking-Unruh radiation is highly sensitive (at ultra low energies) to level repulsion in the chaotic spectrum of the underlying quantum black hole.

We made this explicit by probing a massless scalar field coupled to JT gravity using an Unruh-DeWitt detector. The calculation involves including Euclidean wormhole corrections to a massless scalar bulk two-point function in JT gravity. Due to random matrix universality for quantum black holes, we expect our conclusion to be quite universal and to qualitatively hold in any number of dimensions. One immediate way to test the universality of these ideas is to consider an Unruh-DeWitt detector in super JT gravity or to consider charged versions of JT gravity. Results in this direction are forthcoming.
\\~\\
We end this work by emphasizing three features of our setup that deserve further study. 
\\~\\
\textbf{\emph{Gravitational dressings}}
\\~\\
One important aspect of this work is navigating through different types of operator dressings and operator ordering ambiguities. In particular, for the gravitationally dressed matter modes $A_\omega[f]$ in \eqref{creabn}, we obtained the expectation value of their correlator \eqref{35}. Schematically, we way write this in a canonical language as
\begin{equation}
    [A_{\omega_1}^\dagger[f], A_{\omega_2}[f]]=\delta(\omega_1-\omega_2)+ \text{gravitational corrections}.
\end{equation}
The first contribution is due to the canonical oscillator algebra of the undressed matter modes $a_\omega$. The second contribution is due to canonical commutators of the gravitational variables. Indeed, the matter modes $A_\omega[f]$ include a dressing with gravitational variables. See for example \cite{Donnelly:2015hta,gid3} and section 5 of \cite{ads2}. This should be contrasted with the behavior of the dressed matter modes $B_\omega[f]$ in \eqref{313} for which we write by construction
\begin{equation}
    [B_{\omega_1}^\dagger[f], B_{\omega_2}[f]]=\delta(\omega_1-\omega_2).
\end{equation}
In defining $B_\omega[f]^\dagger B_\omega[f]$ via \eqref{planckp} we have specified a symmetric gravitational dressing for this combination of modes, aimed in a precise way such that the canonical algebra of the modes would receive no gravitational corrections. This means a single mode $B_\omega[f]$ is by itself not a diff-invariant observable but the gravitational dressed composite operator $N_\omega[f] = B_\omega[f]^\dagger B_\omega[f]$ is. On the other hand, the modes $A_\omega[f]$ are well defined diff-invariant operators. The definition of the Unruh-DeWitt detector \eqref{216} requires we use such individually well-defined diff-invariant modes.
\\~\\
More broadly speaking there are numerous operator ordering ambiguities whenever we are promoting a correlator or a field in semiclassical physics to an operator in quantum gravity. One might refer to all of these as choices associated with gravitational dressings \cite{Donnelly:2015hta,gid3}. However in the real world we are all gravitationally dressed composite objects with an implicit choice of dressing. This begs the questions: which dressings are actually natural in quantum gravity and which are just mathematical curiosities? Given a particular physical context, what is the appropriate dressing?

In our particular case of Hawking-Unruh radiation, we were able to pinpoint a natural type of dressing for two distinct experiments. For the Unruh-DeWitt detector experiment we were led to work with the dressed matter modes $A_\omega[f]$ whereas in an experiment which measures the spectral energy density in the heat bath the modes $B_\omega[f]$ turned out to be relevant. Nevertheless, these dressing ambiguities remain largely elusive.
\\~\\
\emph{\textbf{Higher genus bulk correlators}}
\\~\\
Another important aspect of our story is a prescription for how to include Euclidean wormhole corrections to inherently Lorentzian bulk observables in JT gravity. Let us summarize the general idea by an algorithm 
\begin{enumerate}
    \item Find an expression for the bulk matter correlator in the reparameterized metrics \eqref{189}.
    \item Attempt to rewrite that expression as a combination of Schwarzian bilocal operators. An algorithmic way of doing so is to reverse engineer bulk reconstruction in each of the metrics \eqref{189}. Write each term as a Euclidean JT gravity path integral on the disk including corresponding boundary to boundary matter propagators. 
    \item Include Euclidean wormhole contributions to each such individual JT gravity boundary correlator as done in e.g. \cite{phil}. The result is a correlation function in a double-scaled matrix integral.
This is identical to summing over all Riemann surfaces which end on the union of the bilocal lines and the boundary.\footnote{This is true in a precise sense for the two-point function considered in this work. However it is a slight oversimplification for higher point functions. For example in an out-of-time-ordered disk four-point functions, two bilocals cross. This crossing can be avoided on higher genus surfaces if one bilocal traverses a handle whilst the other travels underneath it. In the end however such contributions can also be written in terms of multi-level correlators $\rho(E_1\dots E_n)$ \cite{wophilbert}.} Effectively this leads to replacing $\rho_0(E_1)\dots \rho_0(E_n)$ with $\rho(E_1\dots E_n)$ in the Schwarzian correlators. The correlators $\rho(E_1\dots E_n)$ are multi-level correlators of a double-scaled matrix integral and may be obtained via universal random matrix cluster functions \cite{mehta,paper6}.
    \item Sum over all these boundary correlators in order to obtain the bulk correlator.
\end{enumerate}
One pragmatic way to argue for this prescription is that the resulting bulk correlators end up showcasing certain generic physical principles. One example is level repulsion in Hawking-Unruh radiation as discussed in this work. A second example is the behavior of bulk matter correlators at large distances in a finite entropy system \cite{wophilbert}.
\\~\\
\textbf{\emph{Charged black holes}}
\\~\\
Suppose the level statistics of the black hole is described by random matrix theory with a global symmetry. This would correspond in the bulk to a black hole that carries a conserved (possibly non-abelian) charge. In this case, the level statistics is chaotic in each fixed charge sector \cite{mehta,haake,Kapec:2019ecr}.\footnote{There is a degeneracy of each energy level as the dimension of the representation in the non-abelian case.} This means \eqref{51} has the following schematic replacement:
\begin{align}
\label{chargen}
\rho(M,M-\omega) &= \left\langle \rho(M)\right\rangle \left\langle \rho(M-\omega)\right\rangle + \left\langle \rho(M)\rho(M-\omega)\right\rangle_c  \nonumber \\
&\to \,\, \left\langle \rho_{Q_1}(M)\right\rangle \left\langle \rho_{Q_2}(M-\omega)\right\rangle + \left\langle \rho_{Q_1}(M) \rho_{Q_2}(M-\omega\right\rangle_c,
\end{align}
where the connected piece in the random matrix pair density correlator has a factor $\delta_{Q_1,Q_2}$ and hence only exists if the two conserved charges $Q_1$ and $Q_2$ of the black hole energy levels match. This means that only for this case, one finds level repulsion in the emission spectrum, whereas otherwise one only finds the disk (Schwarzian) gravitational corrections. It would be interesting to investigate this further.
\\~\\
\textbf{\emph{Implications for evaporation?}}
\\~\\
 Our discussion on the Unruh-DeWitt detector experiment is an idealization of a more realistic experiment where the measurement takes place over an infinite amount of time. In a more realistic experiment, we would measure for a finite time $\T$. This can be implemented in the formulas by introducing in the coupling \eqref{26} an additional switching function $\chi(t)$ which has a width of order $\T$. This introduces a factor $\chi(t_1)\chi(t_2)$ in the integrand on the second line of \eqref{29}. If we denote the Fourier transform of $\chi(t)$ by $\hat{\chi}(\omega)$ then \eqref{51} is replaced by a convolution of the previous answer with the frequency content of the switching function
\begin{equation}
R(\omega) \sim \frac{1}{\T} \int_{\mathcal{C}} d\tilde{\omega} \, \hat{\chi}^2(\tilde{\omega}-\omega) \, \frac{\rho(M,M - \tilde{\omega})}{\rho(M)}  \left|\mathcal{O}^1_{M \, M - \tilde{\omega}}\right|^2\,  \frac{\sin^2 z \tilde{\omega}}{\tilde{\omega}^2} \, .
\end{equation}
This convolution replaces \eqref{210} but also for example \eqref{55} with a version that is smooth on frequency scale of order $1/{\T}$.

This implies we must measure for a time ${\T}\gg 1/\beta$ in order to resolve the semiclassical Planckian black body law. The level repulsion in \eqref{51} and the delta spikes in \eqref{55} can only be resolved if we measure for a time ${\T} \gg e^{S_0}$. In a non-evaporating setup this remains a sensible experiment. However for an evaporating black hole on these time scales we are in the regime where the Page curve is decreasing \cite{rw1} and any eternal approximation no longer applies. Nevertheless it is quite natural that the information about the microstates in our eternal model of quantum gravity can only be accessed from a measurement that takes longer than what would be the Page time in an evaporating setup.
    
At the very least this work emphasizes that we should expect extraordinary nonperturbative effects in gravity to highly affect Hawking radiation at long time scales ${\T}\gg e^{S_0}$. These effects can be probed directly in the quantum gravity bulk using an Unruh-DeWitt detector.
\section*{Acknowledgements}
We thank Jordan Cotler, Julius Engelsoy and Joaquin Turiaci for useful discussions. AB and TM gratefully acknowledge financial support from FWO Vlaanderen.
\appendix

\section{Detectors with other boundary conditions}
\label{app:obc}
In section \ref{s:udw} we studied an Unruh-DeWitt detector that couples to a massless scalar field with Dirichlet boundary conditions \eqref{modex}. One might be interested in bosonic bulk matter with other boundary conditions. The result \eqref{237} generalizes as follows. For example, imposing Neumann boundary conditions one finds
\begin{align}
&\average{\Phi[f\rvert u_1,v_1]\Phi[f\rvert u_2,v_2]}_{\CFT}\nonumber\\&\quad=-\frac{1}{4\pi}\ln (F(u_1)-F(u_2))(F(v_1)-F(v_2))(F(v_1)-F(u_2))(F(u_2)-F(v_2))\, .\label{a1}
\end{align}
As a Schwarzian insertion, notice that this operator is not SL$(2,\mathbb{R})$ invariant. For the Fourier transform, we can integrate by parts twice:
\begin{align}
\nonumber &\lim_{{\T} \to +\infty}\frac{1}{\T} \int_{-\T}^{+\T}dt_1\int_{-\T}^{+\T} dt_2\, e^{-i\omega (t_1-t_2)}\,\ln (F(u_1)-F(u_2))\\&\qquad =\frac{1}{\omega^2}\lim_{{\T} \to +\infty}\frac{1}{\T} \int_{-\T}^{+\T}dt_1\int_{-\T}^{+\T} dt_2\, e^{-i\omega (t_1-t_2)}\,\frac{F'(u_1)F'(u_2)}{(F(u_1)-F(u_2))^2} \, .
\end{align}
Notice that hence by integrating the above operator \eqref{a1}, the result \emph{is} SL$(2,\mathbb{R})$ invariant. The full answer \eqref{a1} is a sum over four such terms, each with a generically different phase factor. Summing those phase factors given a greybody factor $\cos^2 z\omega / \omega^2$ replacing the $\sin^2 z\omega / \omega^2$ in the Dirichlet case. Including Euclidean wormhole corrections, one obtains
\begin{equation}
    R(\omega) = 2\frac{\cos^2 z\omega}{\omega^2}\,  \frac{\rho(M,M-\omega)}{\rho(M)}\, \rvert \mo_{M M-\omega}^1\rvert^2\, .\label{a4}
\end{equation}
In terms of an experimental response the effect in this case is quite different to \eqref{234}. The semiclassical answer now blows up for $\omega\ll 1$ due to the double pole in the interference factor. Level repulsion regulates this pole and results in a finite answer for $\omega\ll e^{-S_0}$. For generic conformally invariant boundary conditions, the greybody factor is
\begin{equation}
\frac{1-a \cos 2\omega z}{\omega^2}\,.
\end{equation}
Hera $a=1$ is Dirichlet, $a=-1$ is Neumann, and $a=0$ corresponds to transparant boundary conditions, where one only takes the first two factors in the logarithm of \eqref{a1}. So we see that the behavior of \eqref{234} for $\omega\ll e^{-S_0}$ is an exception to the general rule where the semiclassical answer has a double pole for $\omega\ll 1$ which is regulated by level repulsion.

\section{Airy model and zero energy reference term}
\label{s:zeroref}
Let us briefly address the zero energy \Poincare reference contribution to the spectral energy density \eqref{323}
\begin{equation}
    \bra{0}\omega N[f|\omega]\ket{0}=\frac{V}{4\pi}\,\frac{\rho(0,\omega)}{\rho(0)}\Gamma(1 \pm i\sqrt{\omega})^2+\frac{V}{4\pi}\,\frac{\rho(0,-\omega)}{\rho(0)}\Gamma(1 \pm i\sqrt{\omega})^2 \, .\label{b1}
\end{equation}
We note that the contribution from the second term in \eqref{b1} is essentially negligible for every $\omega$ as it is evaluated in the forbidden region where there is barely any density. \\
Consider first $\rho(0,\omega)$ for $\omega\ll 1$. Both energies are close to the spectral edge and therefore we must use the two-level spectral density of the Airy model instead of \eqref{49}:
\begin{equation}
    \frac{\rho(E_1,E_2)}{\rho(E_2)}=\rho(E_1)+\delta(E_1-E_2)-\frac{K(E_1,E_2)^2}{\rho(E_2)} \, .
\end{equation}
Here the spectral density is
\begin{equation}
 \rho(E) = e^{\frac{2S_0}{3}}\,\Ai'(- e^{\frac{2S_0}{3}} E)^2-e^{\frac{2S_0}{3}}\xi \Ai(- e^{\frac{2S_0}{3}} E)^2\,.\label{b2}
\end{equation}
Furthermore the Airy kernel is\footnote{For a recent derivation of this kernel from a brane correlator computation, see appendix A.4 of \cite{paper6}.}
\begin{equation}
\label{airykernel}
K(E_1,E_2) = \frac{\Ai'(-e^{\frac{2S_0}{3}}E_1)\Ai(-e^{\frac{2S_0}{3}}E_2) - \Ai'(-e^{\frac{2S_0}{3}}E_2)\Ai(-e^{\frac{2S_0}{3}}E_1)}{E_1-E_2}\, .
\end{equation}
This kernel is only relevant for $E_1-E_2$ of order $e^{-S_0}$. Otherwise one finds:
\begin{equation}
    \frac{\rho(0,\omega)}{\rho(0)}=\rho(\omega)\quad,\quad \omega \gg e^{-S_0}\, .
\end{equation}
Using this one finds:
\begin{equation}
    \bra{0}N_\omega[f]\ket{0}=\frac{V}{4\pi}\quad,\quad \omega\gg 1 \, .\label{b6}
\end{equation}
This zero-energy subtraction ensures that the relative spectral energy density $\bra{M}\omega N_\omega[f]\ket{M}$ in \eqref{321} goes to zero smoothly for $\omega \gg M$. 

\begin{figure}[H]
\centering
\includegraphics[width=0.95\textwidth]{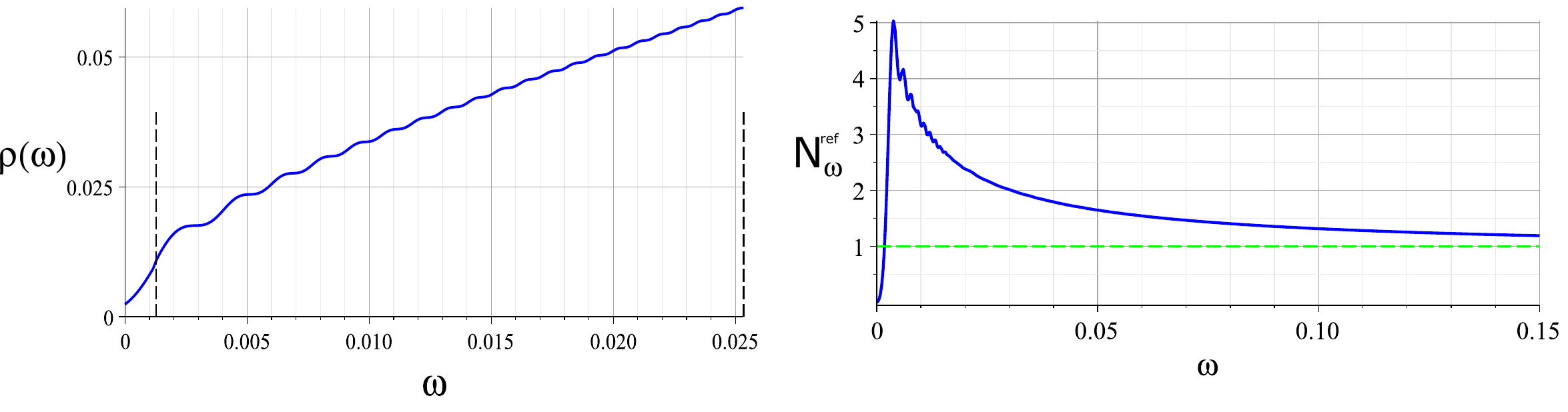}
\caption{
(left) Spectral density $\rho(\omega)$ in JT gravity for any positive energy $\omega$. The Airy answer is accurate for $\omega$ much smaller then the right dotted line. The JT gravity answer \eqref{b7} of \cite{sss2} is accurate for $\omega$ much greater then the first dotted line. (right) $\bra{0}N_\omega[f]\ket{0}$ for any positive $\omega$. The asymptotic limit \eqref{b6} is shown (green). 
}
\label{Urandomref}
\end{figure}

 We furthermore note that \eqref{323} is positive definite for any $M>0$. This is explicit in figure \ref{Urandomz2}. For $M\gg 1$ it is also quite easy to check this explicitly. This whole zero energy contribution \eqref{b1} is strongly suppressed by exponentials of $M$ at any $\omega$ scale.
\\~\\
In order to construct a plot of \eqref{b1} we use the Airy result \eqref{b2} for $\omega\ll 1$, and the smooth JT gravity spectral density $\rho(\omega)$ as determined in \cite{sss2}
\begin{equation}
    \rho(\omega)=\rho_0(\omega)-\frac{e^{-S_0}}{4\pi\omega}\,\cos(2\pi e^{S_0}\int_0^\omega d E\,\rho_0(E)) \, .\label{b7}
\end{equation}
This can be trusted for $\omega\gg e^{-2S_0/3}$. Since $S_0 \gg 1$, these regions overlap so we can construct the whole plot. Both $\rho(\omega)$ and $\bra{0}N_\omega[f]\ket{0}$ are plotted for any positive $\omega$ in figure \ref{Urandomref} which uses $S_0=10$. Notice the wiggles in the latter which shine through in figure \ref{Urandomz2} (right).


\end{document}